\documentclass[12pt]{article}

\input epsf
\def\beq{\begin{eqnarray}}
\def\eeq{\end{eqnarray}}
\def\m{M_*}
\def\msm{M_{\rm SM}}
\def\E{\cal E}
\def\mpl{M_{\rm Pl}}
\def\e{{\epsilon}}

\def\d{\delta^{(N)}(y)}

\def\g{{\cal G}}
\def\lsim{\mathrel{\rlap{\lower3pt\hbox{\hskip0pt$\sim$}}
     \raise1pt\hbox{$<$}}}         %less than or approx. symbol
\def\gsim{\mathrel{\rlap{\lower4pt\hbox{\hskip1pt$\sim$}}
     \raise1pt\hbox{$>$}}}         %greater than or approx. symbol

\begin{document}

\begin{titlepage}

\begin{flushright}
{ NYU-TH-04/08/05}
\end{flushright}
\vskip 0.9cm

\centerline{\Large \bf Looking At The Cosmological Constant}
\vspace{0.4cm}
\centerline {\Large \bf   From Infinite--Volume Bulk}

\vskip 0.7cm
\centerline{\large Gregory Gabadadze}
\vskip 0.3cm
\centerline{\em Center for Cosmology and Particle Physics}
\centerline{\em Department of Physics, New York University, New York, 
NY, 10003, USA}

\vskip 1.9cm

\begin{abstract}

I briefly review the arguments why the braneworld models with 
infinite-volume extra dimensions could solve the cosmological 
constant problem, evading  Weinberg's no-go theorem. Then 
I discuss in detail the established properties of these models,
as well as the features which should be studied further in order to 
conclude whether these models can truly solve the problem. 
This article is dedicated to the memory of Ian Kogan.

\end{abstract}

\vspace{3cm}

\begin{center}

{\it To appear in Ian Kogan Memorial Volume, ``From Fields to Strings:
Circumnavigating Theoretical Physics'', M. Shifman, A. Vainshtein, 
and J. Wheater, eds. (World Scientific, 2004).}

\end{center}

\end{titlepage}

\newpage

\tableofcontents
\newpage

\setcounter{equation}{0}

\section{Two puzzles: cosmological constant  and cosmic coincidence}

Perhaps, the least understood problem of particle physics,
gravity and cosmology is that of the {\it Cosmological Constant}.
The problem stems from a huge mismatch  between  observational  
data and  theoretical expectations. The problem can be  
briefly outlined as follows. The structure in the Universe
(galaxies, clusters etc.) could have formed only 
if the acceleration rate  of the expansion $H_\Lambda $ is 
less than the number that roughly equals to the 
present-day value of the Hubble parameter 
$H_0$ \cite{WeinbergPRL},
\beq
H_\Lambda\,< H_0\,\sim\,10^{-33}\,{\rm eV}\, .
\label{Hexp}
\eeq
Recent observations \cite {cc} appear to confirm a nonvanishing
value of $H_\Lambda$ which nearly saturates the upper bound in 
Eq.~(\ref {Hexp}).
On the other hand, in general relativity (GR) $H_\Lambda^2$ determines the 
scalar curvature of space-time and is related to the vacuum energy density
${\mathcal E}$ as follows:
\beq
\mpl^2 \,H_\Lambda^2\, \sim \,{\mathcal E}\,.
\label{4DH}
\eeq
Here $\mpl\sim 10^{19}~{\rm GeV}$ is the Planck mass.
Moreover, a {\it natural} value of ${\mathcal E}$ due to 
zero-point oscillation energies of known elementary particles can be estimated as ${\mathcal E}\,\gsim\,({\rm TeV} )^4$.
Substituting this value of ${\mathcal E}$ into (\ref {4DH}), one  finds
$H_\Lambda \,\gsim\, 10^{-3}\,{\rm eV}$, which is grossly inconsistent
with  (\ref {Hexp}). This is the essence of the cosmological constant problem 
(CCP).\footnote{We only discuss positive ${\mathcal E}$, all the essential 
arguments apply to negative ${\mathcal E}$ too.}

Historically, the CCP was formulated long before the 
discovery of the cosmic acceleration. One of the first published works 
on the subject was  by  Zel'dovich \cite {Zeldovich} in 1967
where he estimated the contribution of zero-point oscillation energies 
of nucleons to the vacuum energy density. Naturally, 
he found 
$${\mathcal E}\,\sim \,M_{\rm nucleons}^4\,\sim\,({\rm GeV})^4$$
which already gives a result   grossly  inconsistent with  
(\ref {Hexp}).\footnote{According to   colleagues who 
witnessed those developments Zel'dovich got so frustrated by this 
problem that he practically stopped 
doing particle physics and turned his attention  to astrophysics where he 
has made great contributions in the 1970s. 
I thank Misha Shifman for his recollections of that period.} 

The problem only worsened in the 1970s and '80s as  
particle physics made huge steps forward in understanding   Nature 
at exceedingly shorter distances. This only increased
the value of the maximal momentum accessible to a particle 
whose zero-point oscillation energy contributes to the energy density of 
the Universe. Therefore,   estimate of a 
{\it natural} value of ${\mathcal E}$  grew up to  
${\mathcal E}\,\gsim\,({\rm TeV} )^4$. Since then, 
theorists continuously  worked on the problem.  Although no satisfactory 
solution has been found,  nevertheless, as it usually happens, 
many  new useful theoretical aspects   got  uncovered 
in the search process (for a review see, e.g., \cite {Weinberg}).

A dramatic reshaping of the subject took place
with the discovery of the cosmic acceleration at the 
end of the previous millennium \cite {cc}. This discovery triggered 
a fresh tremendous interest in the problem and motivated 
recent developments. So far the discovery 
only sharpened the status of the problem,    making us to 
realize that there are two puzzles that  we have to face. 
In a {\it conventional}  formulation these puzzles 
can be spelled out  as follows: 

(i)  Why is the vacuum energy in the Universe so 
much {\it smaller} than any reasonable estimate 
that follows from particle theories? 
This is the  ``old" CCP. 

(ii) Why is the vacuum energy in the Universe    
{\it comparable} to matter energy?
Or, do we live in a special epoch when the magnitudes of the  
above quantities roughly {\it coincide}?  This is the so called 
cosmic  coincidence problem (CoCoP).

{\it A priori} one could choose to {\it adjust} by hand the 
{\it renormalized} values of the vacuum energy 
density to be equal to ${\mathcal E} \sim (10^{-3}~{\rm eV})^4 $,
to make it  consistent with   observations. There are many 
classical as well as quantum-mechanical contributions 
to the vacuum energy
such that (a) some of  these  contributions 
are many orders of magnitude larger  than  ${\mathcal E} 
\sim (10^{-3}~{\rm eV})^4 $
and  (b) some  of these contributions differ from each other 
by many orders of magnitude. Hence, this adjustment requires an incredible
{\it fine-tuning} of the parameters. Adopting the fine-tuning, we 
could successfully parametrize the observed cosmological 
evolution of the Universe.  However, the fundamental 
questions (i) and (ii) would still remain open since it is not clear
why such different contributions to the vacuum energy had to  
cancel to such an extraordinary accuracy and why the result  
of that extraordinary cancellation should be of the same order as 
the present-day value of the matter density in the Universe. 

In spite of  numerous attempts, neither of the above  puzzles have 
satisfactory explanations so far.\footnote{We are interested in an explanation 
in terms of a low-energy theory. The anthropic approach 
(for a review see, e.g., \cite{anthropic}) seems to  give answers to both 
(i) and (ii), and is certainly a logical possibility. However, 
this is an orthogonal  approach.  
Another logical possibility is that the problem can never  
be understood in terms of low-energy dynamics and is only solved 
due to very contrived effects of ultraviolet physics, which, 
in fact, might not be as contrived as it might seem,  because of symmetries 
of string theory \cite{Dienes}.} Most of the approaches that have been 
developed to solve (i) are  disfavored by  a general no-go 
theorem formulated by S. Weinberg \cite {Weinberg}. As to the solution 
of  (ii), it seems more reasonable to think about it only in 
the context of (i). 

Let us point out that the formulation of the 
question (i) itself contains a loophole which might be 
suggestive of a new approach to the solution of   CCP.
Indeed,  we have no direct experimental way to measure $\E$. 
Instead, we measure space-time curvature through
cosmological observations, and
then determine $\E$ through the Einstein equations.
Thus, claiming that $\E$ should be small we implicitly assume
that the Einstein equations are  valid for arbitrarily large
length scales. This assumption may or may not be correct.
This suggest an alternative approach where $\E$ keeps its natural value
$ \E\gsim $ TeV$^4$, but laws of gravity are modified so that 
large vacuum energy density does not give rise to large space-time 
curvature. Since the discrepancy between the theory and experiment 
manifests itself   at 
enormous distances $\sim H_0^{-1}\simeq 10^{28}$ cm
(i.e., at extremely low energies), to address   CCP it is natural to modify gravity in the  {\it infrared} domain (IR).

Construction of such models was motivated by  the advent 
of the braneworld paradigm, where the standard-model 
fields are localized on a brane while gravity propagates in the 
bulk \cite {ADD} 
(for earlier models see \cite{Akama,RubSh,mishagia}; recent reviews  
can be found in Refs.~\cite{Dick,rev1,rev2}). 
However, making   a consistent theory  of the 
IR-modified gravity became possible only in   models with 
{\em infinite-volume} extra dimensions \cite {DGP,DG}, 
where gravity on the brane transforms from   
four-dimensional to higher-dimensional at very large distances. 
Historically the first was a proposal of Ref.~\cite{Ian}, which gives a 
brane-world realization of a massless and massive gravity.
This was followed by an early  proposal of a theory of 
a metastable graviton \cite {GRS}. However, the latter turned out to be 
an internally inconsistent theory \cite {DGP2,Pilo}.

The  model of Ref. \cite {DGP}, and its higher-dimensional 
generalizations \cite {DG},  paved the way to 
new possibilities of addressing the cosmological 
constant problem through IR modification of gravity, where the vacuum energy (the brane tension)
mostly curves the bulk, while ordinary gravity is trapped on the brane at
observable distances by the
presence of a large Einstein--Hilbert action localized on the brane.

A specific proposal along these lines
was worked out in Ref. \cite{DGS}, where it is argued that the
graviton propagator is modified in the
infrared in such a way that
large wavelength sources, such as the vacuum  energy, gravitate very weakly.
As a result, even a huge vacuum energy does not curve our space.
On the other hand, short wavelength sources, such as
planets, stars, galaxies and clusters gravitate
(almost) normally. The four-dimensional (4D) nonlocal counterpart with similar properties was proposed in Ref.~\cite {ADDG}. 

We will discuss  the framework of Refs.
\cite {DGP,DG} where gravity in general,
and the Friedmann equation, in particular,
are modified for  wavelengths larger than a certain critical value.
This setup can evade the Weinberg no-go theorem.
The cosmological constant problem could  then be remedied
in the following way: Due to the large-distance modification of gravity
the energy density ${\mathcal E} \gsim (1 \,\, {\rm TeV})^4$  does 
not curve the space as it would do in the conventional Einstein gravity.
Therefore, the observed space-time curvature is
small, despite the fact that $\mathcal E$ is huge (as it comes out naturally).
This is the most crucial point of the  approach of 
Refs.~\cite {DGP,DG} -- the point where we depart from the previous 
investigations. Although, as we will see, it is still premature to say
whether this approach leads to a final solution of   CCP,
nevertheless, it seems that all necessary   ingredients  
are present in the model. Future detailed calculations 
will show  whether or not this development is successful. 

Before delving in the issue we would like to mention that the idea of solving 
the cosmological constant problem in theories with extra dimensions and 
branes has a long history (for earliest works
see, e.g.,  \cite {Wett,RS}). However, because of   Weinberg's theorem, 
the solution is only possible   in theories where the extra 
dimensions have infinite (or practically infinite) volume.
Why is this so? A brief answer will be presented below (more
complete discussions are given in Ref. \cite {DGS}).

Recall that if there is a vacuum energy density 
${\mathcal E} \, \ge \,{\rm TeV}^4$ 
in a conventional 4D theory then it unavoidably gives rise to 
the scalar curvature $R\sim H_\Lambda^2$ determined by
(\ref {4DH}).  The vacuum energy density ${\mathcal E} $
is a source of gravity, and, as such, it has to 
curve the space; the only space in   4D theories is
the space in which we live. Hence, our space is curved according to 
(\ref {4DH}), and this is inconsistent with  data. 
However, if there are more than four dimensions,
${\mathcal E}$ could curve extra dimensions
instead of curving  our 4D space \cite {Wett,RS}. 
Consider the following $(4+N)$-dimensional interval: 
\beq
ds^2\,=\,A^2(y)\,{g}_{\mu\nu}(x)\,dx^\mu dx^\nu\,
-\,B^2(y)\,dy^2\,-
\,C^2(y)\,y^2\,d \Omega^2_{N-1}\,,
\label{interval0}
\eeq
where $\mu,\nu =0,1,2,3$, are the indices denoting our 4D world, 
while $y \equiv \sqrt{y_1^2+...+y_N^2}$, 
and $y_n$'s  denote   extra coordinates. What we measure 
in our 4D world is
the curvature 
invariants of the   metric ${g}_{\mu\nu}(x)$. There can exist solutions 
to the  $(4+N)$-dimensional Einstein equations 
in the form of (\ref {interval0})
where ${\mathcal E}$ affects strongly the extra space, i.e., 
the functions $A,B$ and $C$, while 
leaving our 4D space almost intact, with the 4D metric
${g}_{\mu\nu}(x)$ remaining almost flat.

In this case the energy density ${\mathcal E}$ ``is spent'' totally 
on curving up the extra space rather than on curving 
our 4D space. The simplest example of this type is a 
3-brane in six-dimensional space (a local cosmic string)
in which case the tension of the brane is spent on
creating a deficit angle in the bulk, while the brane 
world-volume remains flat (for a discussion see \cite{Sundrum}). 

Such a brane could   be  a good place for our 4D world to  live. 
If one could only obtain the laws of 4D gravity on a brane in 
this setup, this would  be considered as a solution of   CCP
that takes into account all   classical  
and quantum contributions to the cosmological constant!

The same arguments would  apply to higher codimensions. 
Therefore, the paramount goal is to find a mechanism that 
would enable one to obtain 4D gravity on the brane 
embedded in infinite-volume bulk.\footnote{Any compactification of the above setup with the 
compactification radii   smaller than $H_0^{-1}$ would give rise to 
a theory of gravity that flows to the conventional GR in the IR.
The latter would necessarily face Weinberg's no-go theorem,
 for details see \cite {DGS}.} 

The remainder of this article describes
a method of obtaining 4D gravity on a brane in infinite-volume 
extra space.  First, in Sect.~2 we formulate a basic model   
\cite {DGP,DG}. Then  we discuss how   this model evades 
Weinberg's theorem. In Sect.~3 we consider in detail this 
model in five dimensions. Although the 5D model is known 
{\it a priori}   to be unfit  to solve  CCP,  nevertheless,
it is instructive to study this situation in detail. Most of the 
intricate properties  of  the 5D 
model are understood, and one can say that the  
model with appropriate boundary conditions represents a consistent 
theory  of a large-distance modification 
of gravity. In Sect.~4 we turn to similar models in more than five dimensions. 
Here the situation is different. We discuss what is known so far about these 
models and  what needs to be done in order to conclude 
whether this approach  can solve   CCP.  Section 5 contains a 
brief summary.

\section{The origin of the model}

In this section we will formulate the model 
which  was introduced in 5D space-time in 
Ref. \cite {DGP} and later generalized to $D\ge 6$ in \cite {DG}.
We closely follow the presentation of Ref.~\cite {DGS}.

Consider  a brane-world model  in a space with
(asymptotically) flat {\it infinite-volume}
$N$ extra dimensions. Assume that all known standard-model (SM) particles are 
localized on the brane and obey the conventional 4D laws 
of gauge interactions up to very high energies,  of the order of 
the GUT scale, for instance. The gravitational
sector, on the other hand, is spread over the whole $(4+N)$-dimensional
space.  The low-energy action of the model is written as 
\begin{eqnarray}
{S} &=&
\m^{2+N}\int\,d^4x\,d^N y \sqrt{{\bar g}}\,{\mathcal R}_{4+N}({\bar g})
\nonumber\\[2mm]
&+&
\int d^4x
\sqrt{g} \left ({\mathcal E}\,+\,M^2_{\rm ind}\,{R}
+{\mathcal L}_{\rm SM}(\Psi, \msm)\right )\,.
\label{actD}
\end{eqnarray}
Let us discuss  various parts and  parameters of the 
action (\ref {actD}).
${\mathcal L}_{\rm SM}$ is the Lagrangian for   particle physics including 
all SM fields $\Psi$.\footnote{For 
notational simplicity   we use the convention 
that the particles physics  theory,   
including any grand unification (GUT), possibly SUSY GUT, or any other extension  of  standard model,  is denoted as SM.} 
The parameter $\msm$ denotes the ultraviolet (UV) cutoff of   SM.  
Up to that scale  SM obeys the conventional 
4D laws. In the present approach $\msm \gg {\rm TeV} \gg \m$.
Moreover, ${\bar g}_{AB}$  stands for a
$(4+N)$-dimensional graviton $(A,B=0,1,2,...,3+N)$, while 
$y_n,~n=4,5,..,4+N$, denote ``perpendicular'' to the brane coordinates.
For simplicity  we do not consider brane
fluctuations~\footnote{This limitation could be readily lifted.
Indeed, including the brane fluctuations
would produce an almost  sterile Nambu--Goldstone boson,
and heavy modes which could manifest themselves only through
generation of an extrinsic curvature term
on the brane.}.  Thus, the induced metric on the brane is given by
\beq
{g}_{\mu\nu}(x)~\equiv~{\bar g}_{\mu\nu}(x, y_n=0)\,.
\label{ind}
\eeq
Since we discard the brane fluctuations,
the brane can be thought, in the 5D case,  
as a boundary of the extra space or
an orbifold fixed point (in that case the Gibbons--Hawking 
surface term is implied in the action hereafter).
The brane tension is denoted by  ${\mathcal E}$.

The first term in (\ref{actD}) is the bulk Einstein--Hilbert  action
for $(4 +N)$-dimensional gravity, with  the fundamental scale $\m$.
The expression in  (\ref{actD}) has to be understood
as an effective low-energy action  valid for graviton momenta
smaller than $\m$. Therefore, in what follows 
we will imply the presence of an infinite number of 
gauge-invariant high-dimensional
bulk operators suppressed by powers of $\m$. 

The second term in
(\ref{actD}) describes the 4D Einstein--Hilbert (EH) term of the induced
metric. This term plays the crucial role. 
It ensures that at observable distances 
on the brane the laws of 4D gravity are reproduced 
in spite of the fact that there is no localized 
zero-mode graviton.  Its coefficient $M_{\rm ind}$ is another 
parameter of the model. Thus,  the  low-energy action as it 
stands is governed by three parameters 
$\m$, $M_{\rm ind}$ and ${\mathcal E}$.
Let us discuss their natural values separately.

The parameter $M_{\rm ind}$ gets induced  by 
SM-particle loops localized on the brane. 
Such corrections are cut-off by the rigidity scale of   SM, 
$\msm$, i.e.,  the scale above which the SM propagators become soft.
In the present approach this scale is taken to be very high, $\gg$ TeV.
In particular, we will take this scale to be comparable with 
the GUT or 4D Planck scale.\footnote{We set  the thickness of the
brane $\Delta$ to be determined by the
SM scale, $\Delta \sim \msm^{-1}\,$. This might seem a bit unnatural
at a first sight, but there are field theory \cite {junctions} as well 
as string theory constructions \cite {Ignatios} of  branes where such a 
``dynamical'' width is possible.} The loops  induce\,\footnote{$M_{\rm ind}$  
can certainly contain as well the tree-level terms if 
these are present in the original action in the first place.
We will not discriminate between these and induced terms.
$M_{\rm ind}$ will be regarded as a parameter that 
stands in (\ref {actD}).}
the Einstein--Hilbert term in (\ref {actD}),
\beq
M_{\rm ind}^2\,\sqrt{g}\,
{R}({g} )\,,
\label{delta2}
\eeq
where the value of the induced constant $M_{\rm ind}$ is
determined by the relation~\cite {Adler,Zee},
$$M_{\rm ind}^2=i\int d^4x \,x^2\, \langle T(x) \,T(0)
\rangle/96 \,.$$ 
The parameter $M_{\rm ind}$ is proportional
to the scale $\msm$ and to the number of the SM particles.\footnote{The scalars and fermions contribute to $M_{\rm ind}$
with positive sign while the gauge fields
with negative sign.}
Since there are about 60 particles in
the Weinberg--Salam  model, and more are expected in  GUT's,
the value of $M_{\rm ind}$ should be somewhat
larger than $\msm$. In fact, below we define the 4D Planck mass 
as being completely determined by $M_{\rm ind}$,
\beq
\mpl\,\equiv \, M_{\rm ind}\,.
\label{mpl11}
\eeq
Thus, the Planck mass is not a fundamental constant in our approach
but rather a derived scale.
We see that the SM loop corrections 
are capable of creating the hierarchy $M_{{\rm ind}}/M_*$, even 
if the initial value of $M_{{\rm ind}}/M_*$ was  not that large.  
This hierarchy does not amount to fine tuning, 
since such a separation of scales is stable 
under quantum corrections. 
Indeed, say,  $\m$ gets renormalized by all
possible bulk quantum gravity loops. 
However, there are no SM particles in the bulk 
the only scale in there is $\m$. 
Therefore, any bulk loop  gets cut-off
at the scale $\m$, as it is the fundamental 
gravity scale. While, as we discussed above,
the brane SM loops are cut-off by the higher scale 
$\msm$, and this gives rise to the huge 
value of $M_{\rm ind}$ on the brane. 

Finally, let us discuss the value of the brane tension 
(brane cosmological constant), and of the bulk cosmological constant. 
To this end, 
we have to specify our assumptions regarding supersymmetry. We assume that
the high-dimensional theory is supersymmetric, and that supersymmetry is
spontaneously broken only on the brane  (such a scenario with a non-BPS 
brane was considered in \cite{mishagia}). 
The absence of breaking of supersymmetry in the bulk is only 
possible due to infinite volume of the extra space;
SUSY breaking is not transmitted
from the brane into the bulk since 
the breaking effects are suppressed by an infinite 
volume factor.\footnote{In general,
local SUSY  in the bulk does not preclude a negative vacuum energy
density of the order of $\m^{4+N}$.
However, the latter can be forbidden by an unbroken
$R$ symmetry in the bulk. Such a symmetry is often provided by 
string theory.}
Then, the  bulk cosmological term can be set to zero, without any
fine-tuning. On the other hand, the natural
value of  ${\mathcal E}$ can be as low as TeV$^4$, since 
the brane tension can be protected above this value by 
${\mathcal N}=1$ supersymmetry (note that 
$M_{\rm ind}$ can only be protected by a conformal invariance 
which we assume is broken at the scale $\msm$). 
All these properties are summarized in Fig. 1.
\begin{figure}
\epsfxsize=8cm
\centerline{\epsfbox{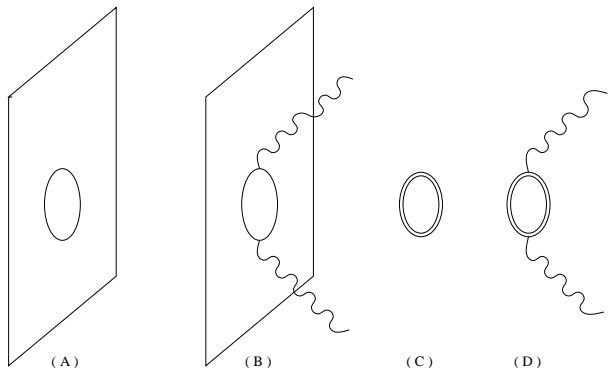}}
\caption{(A) The world-volume 
vacuum diagram of the SM fields 
that renormalizes the brane tension (i.e., the 4D 
cosmological constant). These contributions are protected  
by ${\mathcal N}=1$ world-volume supersymmetry. Therefore, 
they are cutoff by the world-volume SUSY breaking 
scale $M_{\rm susy}$.
(B) The world-volume 
two-point diagram that renormalizes (induces) the EH
term on the brane. These contributions {\it are not protected}
by ${\mathcal N}=1$ supersymmetry. They can only be protected by
conformal invariance which in our model is 
broken at the scale $\msm$ that is close to $\mpl$.
Hence, there can be a hierarchy between (B) and (A).   
(C) The bulk vacuum diagram. Only the bulk particles 
(which do not include the SM particles) are running in this  loop 
(bulk particles are denoted by double lines). 
This diagram is protected by unbroken bulk SUSY. 
Therefore, the cosmological constant in the 
bulk is zero.
(D) The bulk two-point diagram which renormalizes the bulk EH
term. As in (C), only the bulk particles 
are running in the loop. This diagram is cutoff by the bulk 
scale $\m$. Therefore, the natural value of the constant 
in front of the bulk EH term is $\m^{2+N}$; there is huge 
hierarchy between this coefficient and that
of the world-volume EH term coming from (B).}
\label{fig}
\end{figure} 

Let us now turn to the  gravitational dynamics on the brane.
This dynamics is quite peculiar. Despite the fact that the volume
of extra space is infinite, an observer on the brane measures
4D gravitational interaction up to some large cosmologically scales.
The fact that this is so will be studied in detail in the following 
sections.

The no-go arguments discussed in the previous section 
are not applicable to the theories with 
infinite-volume extra dimensions.
The crucial property of this class of theories is 
that despite the unbroken 4D general covariance,
there is no 4D zero-mode graviton.
4D gravity on the brane is mediated by a {\it collective mode}
which  cannot be reduced to any 4D state. 
The fact most important  for us is that the 4D general covariance 
{\it does not} require now all   states to couple universally 
to our ``graviton".   As a result, there is no
universal agent that could mediate supersymmetry breaking from  
SM to all   existing states. Such a situation is impossible
in the finite-volume theories where 4D gravity is mediated by a normalizable
zero mode, which, by   general covariance, must couple
universally and, hence,  mediates supersymmetry breaking.
Moreover, the effect of the brane cosmological term  is to
curve the extra space without inducing a large 4D curvature. 
We stress again that this is  impossible in  finite-volume
theories (i.e., the theories in which the size of the extra space
is smaller than the Hubble size $H_0^{-1}$)
because there the extra components of the metric
are always heavier than $H_0$.

The effective field theory arguments are based on the assumption that
there is a finite number of 4D degrees of freedom below the scale of
the cosmological constant that one wants to neutralize. 
This condition is {\em not}
satisfied in the present model --  it  is a genuinely high-dimensional 
theory in  the far infrared. Therefore, there is an infinite 
number of degrees of freedom
below any nonzero energy scale. As a result, there is no scale below
which   extra dimensions can be integrated out and the theory   reduced to
a  {\it local} 4D field theory with a finite number of degrees of freedom.
In order to rewrite the model at hand as a theory of a single
4D graviton, at any given scale, we have to integrate out an {\it infinite}
number of lighter modes. As    usually happens  in field theory,   
integrating out the light states we get  {\it nonlocal interactions}.
Therefore, the resulting model, rewritten as a theory of a 4D graviton,  
will contain generally-covariant but nonlocal terms. The latter 
dominate the action in the far infrared. 
Of course, in actuality,  the full theory is local -- the apparent 
nonlocality is  an artifact of integrating out light modes. It  
tells us that a local $(4+N)$-dimensional theory can be imitated by a
nonlocal 4D model. The nonlocal terms modify the effective 4D 
equations and neutralize a large cosmological constant.

\section{Five-dimensional model}

In this section we concentrate our attention on a 5D model of 
brane induced gravity -- the so called DGP (Dvali-Gabadadze-Porrati) model \cite{DGP}. 
This model, as a field theory, exhibits many unusual and exciting 
properties that one encounters in theories of large distance 
modified gravity.  In a class of Lorentz-invariant theories 
the model  is right now the only  internally consistent  
theory of large distance modification of gravity (along with
some of its higher-dimensional generalizations, see below).

However, the 5D DGP model cannot solve the ``big'' 
cosmological constant problem because the brane in it has only one 
codimension.  In the best case, the 5D model  can successfully 
parametrize the accelerated Universe  \cite {Cedric,DDG,Landau}
(although the viability of the latter  assertion  still 
needs to be established in greater detail, see below).

Therefore, the 5D DGP model can be regarded as a toy example
on which many intricate features of large distance modified 
gravity can be understood. Some of these features will 
be important in searching  for the theory  of large distance 
modified gravity that could truly solve the ``big'' 
cosmological constant problem.

After reviewing the 5D model in this section, we turn to the 
more general models in $D\ge 6$ in the next section. 
The action of the 5D model  \cite {DGP} is
\beq
S\,=\,{\mpl^2 } \,\int\,d^4x\,\sqrt{g}\,R(g)\,+
{\m^{3} }\,\int \,d^4x\,dy\,\sqrt{{\bar g}}
\,{\mathcal R}_{5}({\bar g})\,,
\label{1}
\eeq
where $R$ and ${\mathcal R}_{5}$ are the four-dimensional
and five-dimensional Ricci scalars, respectively, and
$\m$ stands for the gravitational scale
of the bulk theory. The analog of the graviton mass
is $m_c= 2\m^3/\mpl^2$. The higher-dimensional and
four-dimensional  metric tensors are related as
\beq
{\bar g}(x,y=0)\equiv g(x)\,.
\label{bargg}
\eeq
There is a boundary (a brane) at $y=0$ and ${\bf Z}_2$
symmetry across the boundary is imposed. The presence of the 
boundary Gibbons--Hawking term is implied to warrant  
the correct Einstein equations in the bulk.
Matter fields are assumed to be localized on a 
brane and at low energies, that we observe, 
they do not escape into the bulk. Hence, the matter action 
is completely four-dimensional $S_M=\int d^4x L_M$.
Our conventions are as follows: $ \eta_{AB} = 
{\rm diag}\, [+----]\,; A,B = 0,1,2,3,{\it 5}\,;\mu,\nu = 0,1,2,3.$

\subsection{Perturbative expansion in Newton's constant}

A simplest exercise that tells us a lot 
about the model is to calculate the Green's function 
$D^{\mu\nu;\alpha\beta}$ and the amplitude of interaction 
of two sources $T_{\mu\nu}$ on the brane
\beq
{\mathcal A}_{\rm 1-graviton}\, \equiv 
\,T_{\mu\nu}\,D^{\mu\nu;\alpha\beta}\,
T_{\alpha \beta}\,.
\label{A}
\eeq
In order to perform perturbative calculations one has to fix a  
gauge. One choice, that  was  adopted in \cite {DGP}, is   
harmonic gauge in the bulk $\partial^A h_{AB}= \partial_B h^C_C/2$. 
Then, the momentum-space one-graviton exchange amplitude on the brane 
takes the form:
\beq
{\mathcal A}_{\rm 1-graviton}(p,\,y)\,=\,{T^2_{1/3}\,{\rm exp }\left 
(- p|y| \right ) \over p^2\,+\,m_c\,p}\,,
\label{A13}
\eeq
where we denote the Euclidean four-momentum squared by $p^2$,
\beq
p^2\,\equiv\,-p^{\mu}p_\mu \,=\,-p_0^2\,+\,p_1^2\,+ p_2^2\,+p_3^2\,\equiv
\,p_4^2\,+\,p_1^2\,+ p_2^2\,+p_3^2\,,
\label{psquare}
\eeq
and
\beq
T^2_{1/3} \,\equiv\, 8\,\pi\,G_N \left ( 
T^2_{\mu\nu}\,-\,{1\over 3}\,T\cdot T \right )\,.
\label{T1third}
\eeq
In the expressions above $p$ stands for the square root of $p^2$
\beq
p\,\equiv \,\sqrt{p^2}\,=\,\sqrt{-p_\mu^2}\,.
\label{defpA}
\eeq
The euclidean amplitude (\ref {A13}) was constructed 
by imposing the decreasing  boundary conditions in the 
$y$ direction. 

In principle, one could choose the other sign of the 
square root while solving the equations and obtain the 
euclidean amplitude that grows with $y$
\beq
{\tilde {\mathcal A}}_{\rm 1-graviton}(p,\,y)\,=\,{T^2_{1/3}\,{\rm exp}\left 
( p|y| \right ) \over p^2\,-\,m_c\,p}\,.
\label{A134}
\eeq
The latter  expression differs from the one in (\ref {A13}) 
not only in its $y$ dependence, but also by the 
position of the pole in the denominator.

The above two solutions (\ref {A13}) and (\ref {A134})
are distinguished from each other by the choice of the 
boundary conditions at $y\to \pm \infty$. The choice of the 
decreasing boundary condition in (\ref {A13}) is conventional,
and as we will see below, under this choice one obtains 
the expected  results -- the 4D gravity at 
$r\ll r_c\equiv m_c^{-1}$ is smoothly transitioning to 5D gravity 
at $r\gg r_c$.
On the other hand, the choice of the growing boundary 
conditions in (\ref {A134}) might seem  somewhat unusual.
However, as we will see below, the Minkowski space is 
unstable for this choice, and as a result one obtains 
the so called  selfaccelerated space \cite {Cedric}
which can be used to describe the accelerated expansion of the 
Universe \cite {DDG}.

To reveal these properties  we study the pole structure 
of (\ref {A13}) and (\ref {A134}). Let us  
start with (\ref {A13}). We refer to the branch with 
this choice of the boundary conditions as  the 
``conventional branch'' as opposed to the ``selfaccelerated branch''
specified  by (\ref {A134}). The equation determining the poles on 
the conventional branch is 
\beq
p^2\,+\,m_c\, \sqrt{p^2}\, =\,0\,.
\label{pole1}
\eeq
Hence there are at least two poles, one at $p^2=0$ and 
another one at $p^2=- m_c\,p$. Our goal is to establish where
this poles are located on the complex plane of minkowskian 
momentum square $p_\mu^2$. The transition between euclidean 
momentum square $p^2$ and the minkowskian momentum square 
$p_\mu^2$ is as follows:
\beq
p^2\,=\,e^{-i\pi}\,p_\mu^2\,.
\label{euclmink}
\eeq
Using this we find  poles in minkowskian momentum square
\beq
p_\mu^2=0\,,~~~ p_\mu^2 = m_c^2 e^{-i\pi}\,. 
\label{poles1}
\eeq
As it can be checked, the residue of the $p_\mu^2=0$ pole is zero. 
Therefore, there is no massless mode that can mediate interactions
in this model.  The remaining pole is located on a nonphysical 
Riemann sheet, pointing to a resonance nature of the 
graviton.  The residue in this poles can also be calculated and 
it is positive -- corresponding  to a residues of a positive norm 
state. Hence, on the conventional branch we obtain 
one metastable graviton with the lifetime 
$\tau \sim r_c\sim H^{-1}_0$.

Let us now turn to the ``selfaccelerated branch'' (\ref {A134}).
The poles of this expression are now determined by
\beq
p^2\, -\,\,m_c\, \sqrt{p^2}\,=\,0\,.
\label{pole2}
\eeq
Using the same arguments as above we find the poles,
\beq
p_\mu^2\,=\,0\,,~~~ p_\mu^2\, = \,m_c^2 e^{i\pi}\,. 
\label{poles2}
\eeq
As before, the pole at $p_\mu^2=0$ has zero residue, hence there is 
no massless graviton in this case either. 
On the other hand,  the pole at  $p_\mu^2 = m_c^2 e^{i\pi}$
has a positive residue of a positive norm state. However, this pole 
is located on a physical Riemann sheet. Therefore, 
it describes a tachyon-like  state.  This signals that
the Minkowski space is unstable on this branch. The instability should 
grow with time as
\beq
e^{m_ct}\,.
\label{instability}
\eeq
This is a  welcome feature since this instability could  
signal that the background should be readjusted 
and that the curvature of the new  background  
should be  of the order of $m_c^2$. On the other hand, this 
is roughly the curvature that is needed to describe the 
accelerated universe (for a modern review on theory and observations, 
see \cite {Igor}).

The above perturbative arguments can be generalized to a full-fledged 
nonperturbative analysis by looking at exact cosmological solutions 
of the model \cite {Cedric,DDGL,DDG}. One can 
the exact cosmological equations for studying the evolution on the 
brane. What is important here is the expression for the 
Friedmann equation  on the brane. In terms of the Hubble parameter 
of the 4D brane world-volume $H$, the latter equation takes the form

\vspace{0.1in}
{\it (I) Conventional branch, i.e., decreasing boundary conditions 
at $y\to \pm \infty$ (compare with (\ref {pole1})}):
\beq
H^2\,+\,m_c\,H\,=\,0\,.
\label{F1}
\eeq
There are two solutions to the above equation:

Solution (A)
\beq
H\,=\,0\,.
\label{F1sol1}
\eeq
This solution corresponds to the Minkowski space 
of the conventional branch. Small perturbations about this 
space are stable.  On this solution  the cosmological evolution 
transitions from a 4D regime when $H\gg m_c$ to the 5D regime when
$H\ll m_c$. This behavior might be useful for certain cosmological 
issues, however, it cannot explain the accelerated expansion of the 
Universe.

Solution (B)
\beq
H\,=\,-m_c\,.
\label{F1sol2}
\eeq
This corresponds to a collapsing Universe with the scale factor 
${\rm exp}(-m_ct)$ and the typical time scale determined by 
$r_c$.

\vspace{0.1in}
{\it (II) Selfaccelerated  branch, i.e., increasing  boundary conditions 
at $y\to \pm \infty$ (compare with (\ref {pole2})}):
\beq
H^2\,-\,m_c\,H\,=\,0\,.
\label{F2}
\eeq
These are empty space Friedmann equations. 

Solution A$^\prime$
\beq
H\,=\,0\,.
\label{F2sol1}
\eeq
This is a Minkowski solution of the selfaccelerated branch.
However, as we discussed above, the small perturbations 
about this branch reveal the exponential 
instabilities of the type (\ref {instability}) with the 
typical time scale determined by $r_c$. There are two questions in 
this regards: 

(1) Where this instability leads the theory? 

(2) Whether this instability can be used to mimic the 
accelerated expansion of the Universe? 
These questions were  not studied yet.

Solution B$^\prime$
\beq
H\,=\,m_c\,.
\label{F2sol2}
\eeq
This is a selfaccelerated solution found by Deffayet 
\cite {Cedric}. This solution was shown to describe 
successfully the accelerated expansion of the Universe 
\cite {DDG,Landau}. The question whether this solution itself is stable 
with respect to small fluctuations needs further detailed studies in the light of 
the results of Refs. \cite {Luty,Nicolis} where it was shown that in 
a particular limit of the theory there is a ghost-type excitation on 
the selfaccelerated background (\ref {F2sol2}). 
The question whether this ghost is present on the selfaccelerated
background in the full theory and is not an artifact of the 
particular limit taken in Refs. \cite {Luty,Nicolis} needs to be 
studied.

Some of the discussions presented above 
were based on purely perturbative arguments
(although in all the cases the exact 
results could also be obtained). In this regard, 
it is appropriate to wonder about  the limitations
of the perturbation theory in the present case.
As we will see, it turns out that the naive perturbative
expansion in Newton's constant breaks down unusually 
early as compared to the standard GR.
In  general terms the reason for this breakdown is 
as follows. The 5D model  has two {\it dimensionful} parameters:
the Newton constant $G_N$  and the graviton lifetime 
$m_c$. The naive perturbative expansion in 
powers of $G_N$ is contaminated by powers of $1/m_c$.
Hence, for small values of $m_c$ perturbation theory 
breaks down for the unusually low value of the energy scale.

The reason for the breakdown of perturbation theory at a low scale 
can be traced back to terms in the graviton propagator 
that contain  products of the structure
\beq
{p_\mu \,p_\nu\over m_c p} \,,      
\label{sing}
\eeq
with similar structures or with the flat space metric. 
These terms do not manifest themselves in 
physical amplitudes at the linear level since they are multiplied 
by conserved currents, however,  they enter nonlinear diagrams 
leading to the breakdown of perturbation theory similar to   
massive non-Abelian gauge fields or 
massive gravity.\footnote{Unfortunately, the  massive gravity in 4D 
\cite {PF} is an unstable  theory 
\cite {Deser} with an instability 
time scale that can be rather short \cite {GaGr}.}
However, this breakdown is 
an artifact of an ill-defined perturbative expansion -- 
the  known  exact solutions of the model have no trace of 
the breakdown (see Refs. \cite {Arkady,DDGV}). 
This shows that if one sums up all the tree-level 
perturbative diagrams, then the breakdown problem should disappear. 

For a source of mass $M$ and the Schwarzschild radius $r_M\equiv 
2G_N M$, the perturbative breakdown scale takes the form \cite {DDGV}
\beq
r_* \,\equiv \,(r_M\,r^2_c)^{1/3}\,. 
\label{rstar}
\eeq
This is a scale at which nonlinear interactions in a naive 
perturbative expansion in $G_N$ become comparable with the linear terms 
(below we will discuss  in detail the physical meaning of this scale). 
For a source such as the Sun, the hierarchy of the scales is as follows:
\beq
r_M \sim 3\,10^5~{\rm cm}\,  \ll \,r_*\sim 3\,10^{20}~{\rm cm} \, 
\ll r_c\sim 10^{28}~{\rm cm}\,. 
\label{hierarchy}
\eeq
It is interesting to note that for cosmological solutions 
of the Friedmann--Robertson--Walker (FRW) type $r_*\sim r_c$. 
(The same is true for 
very law energy density  sources.) Therefore, the 
perturbative calculations described above give valid results for the 
FRW type solutions only at distance/time scales larger that $r_c$.
This is confirmed by exact cosmological solutions. 

Hence, the conclusions of the above perturbative calculations can be used
to state that there are two branches of solutions, 
that have different behavior at $t>r_c$. These are
the conventional solution  and the selfaccelerated branch described above. 
The Minkowski space is stable on the  conventional branch, however 
it is unstable on the  selfaccelerated branch. 

\subsection{Constrained perturbation theory}

In this section we consider a possibility of 
modifying the linearized perturbation theory in the DGP model
by introducing certain new terms  that would enable to remove 
the singular in $m_c$  terms from  the propagator.
We closely follow Ref. \cite {GG}.

For this we recall that the the breakdown of the perturbative expansion 
in $G_N$ can be traced back to  the expression for the trace of 
$h_{\mu\nu}\equiv g_{\mu\nu}-\eta_{\mu\nu}$ 
which in the harmonic gauge takes the form 
\cite {DGP}
\beq
{\tilde h}^\mu_\mu(p, y=0) \,=\,-\,{T \over 3\,m_c\,p}\,.
\label{trace}
\eeq
(Hereafter the tilde  denotes Fourier-transformed quantities,
and we put $8\,\pi\,G_N\,=1$.). From this expression we learn 
that: (i) ${\tilde h}^\mu_\mu $ 
is a propagating field in this gauge;
(ii) ${\tilde h}^\mu_\mu $ propagates as a 5D field, 
i.e., it does not see the brane kinetic term; (iii) The 
expression for ${\tilde h}^\mu_\mu $  is singular in the 
limit $m_c\to 0$.  The gauge dependent part of the 
momentum-space propagator ${\tilde D}(p,y)$ 
contains the terms $p_\mu \,p_\nu {\tilde h} $, which, due to 
(\ref {trace}), give rise to the 
singular in $1/m_c$ term. Hence, to understand the origin of the 
breakdown of perturbation theory, one should look at the 
origin of the $1/m_c$ scaling in (\ref {trace}).

The singular behavior of ${\tilde h}^\mu_\mu $  
is a direct consequence of the fact that the four-dimensional 
Ricci curvature $R(g)$  in the linearized approximation 
is forced to be zero  by the  $\{55\}$ and/or 
$\{\mu 5\}$ equations of motion. This can be seen by direct 
calculation of $R$ and of those equations, but  it is more instructive 
to see this by using the Arnowitt--Deser--Misner (ADM) decomposition.  The $\{55\}$ equation reads
\beq
R\,=\,(K^\nu_\nu)^2\,-\,K_{\mu\nu}^2\,,
\label{ADM}
\eeq
where $K_{\mu\nu}$ denotes the extrinsic curvature. Since  
$K\sim {\mathcal O} (h)$ the above equation implies that 
the four-dimensional curvature $R\sim {\mathcal O} (h^2)$ and  in the 
linearized order $R$ vanishes.  Let us now see how this leads to 
the singular behavior of  $h$ in (\ref {trace}). The junction 
condition across the brane  contains  two types of terms: 
there are  terms proportional to $m_c$ and there are terms that 
are  independent of $m_c$. The former come from the bulk Einstein---Hilbert 
action while the latter appear due to the world-volume Einstein--Hilbert 
term. In the trace of the junction condition
the $m_c$ independent term is  proportional to the four-dimensional 
Ricci scalar $R$. On the other hand, as we argued above, $R$ has no 
linear in $h$ term in the weak-field expansion, simply because 
these terms cancel out  due to the  $\{55\}$ and/or 
$\{\mu 5\}$ equations.  Therefore, in the linearized approximation 
the junction condition  contains only the terms that  
come from the bulk. These terms are proportional to 
$m_c h$. This inevitably leads to the trace of $h$ (\ref {trace}) 
that is singular in the $m_c\to 0$ limit and 
triggers the breakdown of the perturbative approach as discussed above.

The above arguments  suggest  that the two limiting procedures, 
first truncating the small $h$ expansion and only then taking  
the $m_c\to 0$ limit, do not commute with each other. 
Therefore, the right way to perform the calculations is either 
to look at exact solutions of 
classical equations of motion, as was argued in 
\cite {Arkady,DDGV}, or to retain at least 
quadratic terms in the equations. The obtained results 
won't be singular in the $m_c\to 0$ limit.

However, neither of the above approaches addresses the 
issue of  quantum gravitational loops. Since the loops can only be 
calculated within a well-defined perturbation theory, 
one needs to construct a new perturbative expansion 
that would make diagrams tractable at short distances.

Below  we will rearrange 
perturbation theory in such a way that the consistent 
answers be obtained in the weak-field approximation.  
This can be achieved if the linearized gauge-fixing terms 
can play  the role similar to the nonlinear terms. 
We will see that this  requires a certain nontrivial   
modification of the linearized theory  and of gauge-fixing 
procedure. 

We recall that in the DGP model the boundary (the brane) 
breaks explicitly  translational invariance in the 
$y$ direction, as well as the rotational symmetry that 
involves the $y$ coordinate.
However, this fact is not reflected in the linearized 
approximation -- the linearized theory that follow from 
(\ref {1}) is invariant under five-dimensional 
reparametrizations.\footnote{If instead of the boundary 
we consider a dynamical brane of a nonzero tension,  
then the five-dimensional Poincare symmetry is nonlinearly 
realized and one has to include a Nambu--Goldstone mode on the brane.}
This line of arguments suggests 
to introduce constraints in the linearized theory that would account
for the broken symmetries.  It is clear that an arbitrary set 
of such constraint cannot be consistent with equations of 
motion with boundary conditions on the brane and at $y\to \infty$. 
However, by trial and error a 
consistent set  of constraints and gauge conditions can be found.
Below we introduce this set of equations step by step.
We start by imposing the following condition:
\beq
B_\mu \, \equiv\, \partial_\mu h_{55} \,+\,
\partial^\alpha h_{\alpha \mu}\,
=\,0 \,.
\label{bmu}
\eeq
Furthermore,  to make the kinetic term for the 
$\{\mu 5\}$ component invertible we set a second condition:
\beq
B_5\,\equiv\,\partial^\mu\,h_{\mu 5}\,=\,0\,.
\label{b5}
\eeq 
At a first sight, the two conditions 
(\ref {bmu}) and  (\ref {b5}) fix all the $x$-dependent gauge 
transformations and make the gauge kinetic terms 
nonsingular and invertible. However, at a closer inspection this 
does not appear to be satisfactory.  
One can look at the $\{\mu\nu \}$ component of the equations of 
motion  and integrate this equation with respect to $y$  
from $-\epsilon $ to $\epsilon$,
with $\epsilon \to 0$. After the integration,  all the terms 
with $B_\mu$ and $B_5$ vanish. The resulting equation (which is just 
the Israel junction condition) taken by its own, is 
invariant under the following four-dimensional transformations
\beq
h^\prime_{\mu\nu}(x^\prime, y)|_{y=0}\,=\,h_{\mu\nu}(x, y)|_{y=0}\,+\,
\partial_\mu \zeta_\nu |_{y=0}\,+\, 
\partial_\nu \zeta_\mu |_{y=0}\,. 
\label{branetransf}
\eeq
This suggests that in the $m_c\to 0$ limit 
the gauge kinetic term on the brane is not invertible.
As a result, the problem of a precocious breakdown of perturbation theory
discussed in the previous section arises. 
To avoid this difficulty one can  introduce the following term 
on the brane world-volume:
\beq
\Delta S\, \equiv \, 
-\, \mpl^2\,\int\,d^4x \,dy \,\delta (y)\,\left ( \partial^\mu h_{\mu\nu} - 
{1\over 2} \partial_\nu h^{\alpha}_{\alpha} \right )^2\,.
\label{gauge}
\eeq
This makes the graviton kinetic term of the brane invertible even in the 
$m_c \to 0$ limit. At this stage, the partition function can 
be {\it defined} as   
\begin{eqnarray}
Z_{\rm gf}&=&{\rm lim}_{\alpha,\gamma\to 0}
\int dh_{AB} \, 
{\rm exp} \Big( i \, S + i \, \Delta S
 \nonumber\\[3mm]
&+& \left. i \,
 \m^3\int d^4x dy 
\, \left \{{B_5^2\over 2\gamma} + {B_\mu^2 \over 2\alpha} 
\right \} \right )\,.
\label{zgf}
\end{eqnarray}
Here $S$ and $\Delta S$  are given in (\ref {1}) and 
(\ref {gauge}) respectively, 
and the limit $\alpha,\gamma\to 0$ enforces 
(\ref {bmu}) and (\ref {b5}). 
Before proceeding  further, notice that 
Eqs. (\ref {bmu}) and (\ref {b5}) would have
been just gauge-fixing conditions 
if the boundary were absent (e.g., in a pure 5D theory with no brane). 
However, in the present case, the above equations,
when combined with the junction condition across the brane, 
enforce certain boundary conditions on the brane. Therefore,  
Eqs. (\ref {bmu}) and (\ref {b5}) do more than 
gauge-fixing, and  $\gamma$ and $\alpha $ 
cannot be regarded as gauge fixing parameters. 
The prescription given by (\ref {zgf}) is to 
calculate first all Green's functions and then take the limit 
$\alpha, \gamma \to 0$.  Because of this, the results of 
the present calculations differ from \cite {DGP} where 
other boundary conditions were implied.

Using (\ref {zgf}) we calculate below 
the propagator $D$ and the amplitude ${\mathcal A}$ defined in (\ref {A}).
We will see that there are no terms in $D$ that blow up as $m_c\to 0$.

We start with the equations of motion that follow from 
(\ref {zgf}). The $\{\mu\nu\}$ equation on the brane reads
\beq
&{m_c\over 2}& \int_{-\epsilon}^{+\epsilon}\,dy \,
\left ( \partial_D^2\,h_{\mu\nu}\,-\,
\eta_{\mu\nu}\,\partial_D^2\,h^\alpha_\alpha  \,
+\,\partial_\mu\partial_5 h_{5\nu}\,+\,
\partial_\nu\partial_5 h_{5\mu}\, - \,2\, \eta_{\mu\nu}
\partial^\alpha \partial_5 h_{5\alpha}\, \right )
\nonumber \\[2mm]
&+& G_{\mu\nu}^{(4)}-
 (\partial_\mu \partial_\alpha h_{\alpha \nu} 
+\partial_\nu \partial_\alpha h_{\alpha \mu} 
- \partial_\mu \partial_\nu h^\alpha_\alpha -\eta_{\mu\nu}
 \partial_\alpha \partial_\beta h^{\alpha \beta} +{1\over 2} 
 \eta_{\mu\nu} \partial_4^2 h^\alpha_\alpha  ) 
 \nonumber \\[2mm]
 &= &T_{\mu\nu}\,,
\label{munubrane}
\eeq 
where 
$$
\epsilon \to 0 ,  \quad
 \partial_D^2\equiv \partial_A\partial^A , \quad
 \partial_4^2\equiv \partial_\mu\partial^\mu
 \,.
 $$
In (\ref {munubrane}) we retained only 
terms that do not vanish in the  $\epsilon \to 0$ limit.
Furthermore, $G_{\mu\nu}^{(4)}$ denotes the 4D Einstein tensor,
\beq
G^{(4)}_{\mu\nu}&=& \partial_4^2\,h_{\mu\nu}- \partial_\mu 
\partial_\alpha 
h^\alpha_\nu  - \partial_\nu  \partial_\alpha 
h^\alpha_\mu + \partial_\mu\,\partial_\nu\,h^\alpha_\alpha
\nonumber \\[2mm]
&-&  \eta_{\mu\nu}
\partial_4^2\,h^\alpha_\alpha +  \eta_{\mu\nu}\,\partial_\alpha
\,\partial_\beta\,h^{\alpha \beta}\,.
\label{G45}
\eeq
The $\{\mu\nu\}$ equation in the bulk takes the form
\beq
& \partial_D^2\,h_{\mu\nu}& -\,\eta_{\mu\nu}\,\partial_D^2\,h^\alpha_\alpha
\,-\, \partial_\mu \partial^\alpha h_{\alpha \nu} \,-\,
\partial_\nu \partial^\alpha h_{\alpha \mu} \,+\,
\partial_\mu \partial_\nu h^\alpha_\alpha \nonumber \\[2mm] 
&+& \eta_{\mu\nu}\,\partial_\alpha \partial_\beta h^{\alpha \beta}\,
+\,\eta_{\mu\nu}\, \partial_4^2\,h_{55}\,-\,\partial_\mu\partial_\nu \,h_{55}
\,+\,\partial_\mu\partial_5 h_{5\nu}\nonumber \\[2mm] 
&+& \partial_\nu\partial_5 h_{5\mu}\, -\,2\, \eta_{\mu\nu}
\partial^\alpha \partial_5 h_{5\alpha}\,-\,{1\over \alpha}\,
( \partial_\mu\partial_\nu h_{55}\,+\, \partial_\mu \partial^\alpha
h_{\alpha \nu})
\nonumber \\[2mm]
&=& 0\,.
\label{munubulk}
\eeq
At the next step we turn to the $\{\mu 5 \}$ equation  which 
can be written as 
\beq
\partial_4^2 h_{\mu 5} \,-\,\partial_\mu \partial^\alpha h_{\alpha 5}
-\partial_5 (\partial^\alpha h_{\alpha \mu} -\partial_\mu h^\alpha_\alpha)
 - {1\over \gamma}\, (\partial_\mu \partial^\alpha h_{\alpha 5})\,=\,0\,.
\label{mu5}
\eeq
Finally, the $\{5 5 \}$ equation takes the form
\beq
\partial^2_4\, h^\alpha_\alpha -\partial_\mu\partial_\nu h^{\mu\nu}
-{1\over \alpha} (\partial^2_4\,h_{55} + \partial_\mu\partial_\nu h^{\mu\nu})
\,=\,0\,.
\label{55}
\eeq
The limit
$\alpha, \gamma \to 0$ should be taken
after the calculation is carried out.

We turn to the momentum space with respect to four world-volume coordinates,
\beq
{h}_{AB}(x, y)\, = \,\int d^4p \,e^{ipx}\, {\tilde h}_{AB}(p, y)\,.
\label{mom}
\eeq
From the above equations we calculate the response of gravity to the source 
$T_{\mu\nu}$. In the  limit $\alpha, \gamma \to 0$ the results are 
\beq
{\tilde h}_{\mu\nu}(p, y)\, \to \, 
{1\over p^2\,+\,m_c\,p}\,\left ( T_{\mu\nu}\,-\,
{1\over 2}\,\eta_{\mu\nu}\,T \,{ p^2\,+\,2\,m_c\,p \over p^2\,+\,
3\,m_c\,p}\right)e^{-p|y|}\,.
\label{hmunu}
\eeq
We note that  this expression is regular in the $m_c\to 0$ limit.
This is contrary to what happens in the 
harmonic gauge \cite {DGP} where   singular terms are present.

For the off-diagonal components we find that 
${\tilde h}_{\alpha 5}\sim \gamma \,p_\alpha $, and 
\beq
{\tilde h}_{\alpha5}(p, y) \to 0\,.
\label{hmu5}
\eeq
Finally,   
\beq
{\tilde h}_{55} \,\to - \,{r\over 2}
{\tilde h}^\alpha_\alpha \,\to  \,
{r\over 2}\,{T\over p^2+3m_cp}\,e^{-p|y|}\,,
\label{h55}
\eeq
with $r\equiv (p^2+2m_c p)/(p^2+m_cp)$.
The amplitude on the brane takes the form
\beq
{\mathcal A}_{\rm 1-graviton}(p, y=0)\,
=\,{1\over p^2\,+\,m_c\,p}\,\left ( T_{\mu\nu}^2\,-\,
{1\over 2}\,T^2 \,{ p^2\,+\,2\,m_c\,p \over p^2\,+\,3\,m_c\,p}\right)\,.
\label{Adgp}
\eeq
A remarkable property of this amplitude is that 
it interpolates between the 4D behavior at $p\gg m_c$
\beq
{\mathcal A}_{4D}(p, y=0)\,
\simeq\,{1\over p^2 }\,\left ( T_{\mu\nu}^2\,-\,
{1\over 2}\,T^2 \right)\,,
\label{A4D}
\eeq
and the 5D amplitude at $p\ll m_c$
\beq
{\mathcal A}_{5D}(p, y=0)\,
\simeq \,{1\over m_c\,p }\,\left ( T_{\mu\nu}^2\,-\,
{1\over 3}\,T^2 \right)\,.
\label{A5D}
\eeq
This amplitude has no van Dam--Veltman--Zakharov  (vDVZ) discontinuity \cite {Iwa,Veltman,Zakharov}.

It is instructive to rewrite the amplitude (\ref {Adgp})  
in the following form:
\beq
{\mathcal A}_{\rm 1-graviton} \,=\,{T^2_{1/2} \over p^2\,+\,m_c\,p}\,
+\,{1\over 6}\,T^2 \,{g(p^2)\over p^2\,+\,m_c\,p}\,,
\label{Adgpsplit}
\eeq
where
\beq
T^2_{1/2} \,\equiv\,\left 
( T^2_{\mu\nu}\,-\,{1\over 2}\,T\cdot T \right )\,,
\label{T1half}
\eeq
and 
\beq
g(p^2)\,\equiv \, {3\,m_c\,p \over p^2\,+\,3\,m_c\,p}\,.
\label{gp}
\eeq
The first term on the right-hand side of (\ref {Adgpsplit}) is 
due to  two transverse polarizations of the graviton, 
while the second term 
is due to an extra scalar polarization. The scalar
acquires a momentum-dependent form-factor. The form-factor is such that 
at subhorizon  distances, i.e., when $p\gg m_c$,  
the scalar decouples. At these scales
the effects  of the extra polarization is suppressed by a factor 
$m_c/p$ (e.g., in the Solar system this is less than $10^{-13}$).
However, the scalar polarization 
kicks in at  superhorizon scales, $p\ll m_c$,  
where the five dimensional laws or gravity are restored. 
\begin{figure}
\centerline{\epsfbox{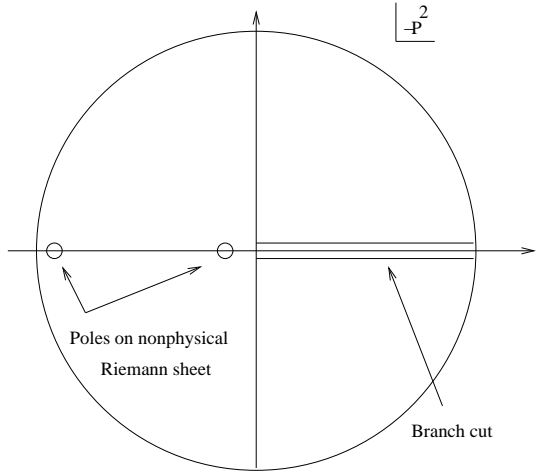}}
\epsfysize=6cm
\vspace{0.1in}
\caption{\small The complex plane of Minkowskian momentum square 
with the branch cut and poles on a second Riemann sheet. The  
physical Riemann sheet is pole free.}
\label{figure}
\end{figure} 

Let us discuss the above results in more detail. 
For this we study the pole structure of the amplitude  
(\ref {Adgpsplit}). There are two nontrivial poles  
\beq
p^2\,=\,-\,m_c\,p\,,~~~~{\rm and}~~~p^2\,=\,-3\,m_c\,p\,.
\label{poles}
\eeq
Let us find the positions of these poles 
on a complex plane of the Minkowski momentum square $p_\mu^2$,
where $p^2=p_\mu^2 {\rm exp}{(-i\pi)}$. For this we note that 
there is a branch cut from zero to plus infinity on the 
complex plane (see Fig. 2). The pole at $p^2=0$ 
is just the origin of the branch cut with zero residue. 
Because of the  cut the 
complex plane has many sheets (the propagator is multivalued 
function due to the square root in it). It is straightforward to show that 
both of the poles in (\ref {poles}) are on the {\it unphysical},
second Riemann sheet. Moreover, the positions of these poles
are far away from the branch cut (usual particle physics resonances
appear on unphysical sheets  close to the branch cut, 
the above poles, however,  are located on a negative semiaxis 
of the second Riemann sheet). Hence, the physical Riemann sheet is pole 
free. The poles on a nonphysical sheet 
correspond to metastable states that do not appear as {\it in} and {\it out}
states in the S-matrix \cite {Veltman}. Using the contour of Fig. 2
that encloses the plane with no poles, and taking 
into account the jump across the cut, the four-dimensional 
K\"allen--Lehmann  representation can be written for the 
amplitude (\ref {Adgpsplit}). The latter warrants four-dimensional 
analyticity, causality and  unitarity of the amplitude 
(\ref {Adgp}). Although the above interpretation is the only correct one,
one could certainly adopt the following provisional picture 
that might be convenient for intuitive thinking. The second pole 
in (\ref {poles}) can be interpreted as a ``metastable ghost'' 
with a momentum-dependent decay width that accompanies the fifth 
polarization and cancels its contributions at short distances. 
Remarkably, this state does 
not give rise to  the usual instabilities because it  can only appear in
{\it intermediate states} in Feynman diagrams, 
but  does not appear in  the {\it in} and {\it out} states in the S-matrix 
elements. In this respect, it is more appropriate to think 
that the scalar graviton  polarization acquires the 
form-factor $g(p)$ (\ref {gp}).

The above results seem somewhat puzzling from the point of view of the 
Kaluza--Klein (KK) decomposition. Conventional intuition would suggest that 
the spectrum of the KK modes consists of massive spin-2 states. 
The K\"allen--Lehmann representation for the 
amplitude as a sum with respect to these  massive states 
would give rise to the tensorial structure  
where  the first term on the right-hand side of 
(\ref {Adgpsplit}) is proportional to $T^2_{1/3}$, 
instead of $T^2_{1/2}$. 
In this case, the remaining part of the amplitude on the right-hand side would have 
a {\it negative} sign. This might be thought of 
as a problem. However, this is not so. The crucial difference 
of the present approach from the conventional KK theories is that 
the effective 4D states are mixed states of an infinite number 
of tensor  and scalar modes. What is responsible for the mixing between  
the different spin states is the brane-induced term and 
the present procedure of imposing the constraints. 
In the covariant gauge that we discuss the 
trace of $h$ propagates and mixes with tensor fields.
From the KK point of view this would look as an infinite 
tower of states with wrong kinetic terms. However, at 
least in the linearized  approximation, the trace is a gauge  artifact.
Nevertheless, the effect of the trace part is 
that the true physical eigenmodes do not carry a definite 
four-dimensional spin of a local four-dimensional theory 
(see also \cite {Romb}).
Because of this  there is no reason to split the amplitude 
(\ref {Adgpsplit}) into the term that is proportional to $T^2_{1/3}$ 
and the rest.

The question of interactions of these states 
in the full  nonlinear theory is not yet understood.  
What happens with the diagrams in which the  ``metastable ghosts'' 
propagate in the  loops (the unitarity cuts of which 
should give production of these multiple states) 
remains unknown. However, since the theory possesses 
4D reparametrization invariance, we expect that these questions will find 
answers similar to those of  non-Abelian gauge fields. Further  
studies are being conducted to understand these issues.

To summarize briefly, in this approach perturbation theory is 
well-formulated. The resulting amplitude interpolates between 
the 4D behavior at 
observable distances and 5D behavior at superhorizon scales.
This is due to the scalar polarization of the graviton 
that acquires a momentum-dependent form-factor. As a result, 
the scalar decouples with high accuracy from the 
observables at subhorizon distances. 

The model can potentially evade the no-go theorem for massive/metastable  
gravity \cite {Veltman}, that states that for the cancellation of the 
extra scalar polarization one should introduce a ghost that would give 
rise to instabilities \cite {Veltman, DGP2}. In the present model, 
at least in the linearized approximation, such instabilities do not occur.
The convenient (although not precise)  
picture is to think of a ``metastable ghost'' that 
exists only as an intermediate state in Feynman diagrams 
which does not appear in the final states at least in the linearized
theory. Since this state cannot be emitted in physical processes, it 
does not give rise to the usual instability. The latter property 
is similar to the observation made in the ``dielectric'' regularization of the 
DGP model in \cite {Romb}. 

The questions that remain open concern the gauge-fixing and interactions 
in the full nonlinear theory where the Faddeev--Popov 
ghosts are expected to play a crucial role. These issues will be 
addressed elsewhere. 

\subsection{Magic of nonlinear dynamics}

Exact static solutions in models of gravity carry a great deal of 
information on the gravitational theories themselves. Hence, finding 
these solutions in models that modify gravity at large distances is an 
important and interesting task. In this section, following Ref.
\cite {GI}, we will study the  Schwarzschild solution  in the 5D DGP 
model \cite {DGP}. It is complicated to find this solution  
since even at distances much larger than the Schwarzschild radius 
of the source, full nonlinear treatment is required 
\cite {DDGV}.  The first approximate 
solution was obtained in Ref. \cite {Andrei} and subsequently by 
the authors of  Refs. \cite {MassimoBH,Lue,Sio,Nicolis}. 
The solution should interpolate between very different distance scales. 
These scales are: the 4D gravitational radius of the source 
of a mass $M$, $$r_M\equiv 2G_N M\,,$$
the large distance crossover scale $r_c\sim 10^{28}$ cm, and an 
intermediate scale, first discovered by Vainshtein in massive 
gravity \cite {Arkady}, which in the DGP model reduces \cite{DDGV} to
\beq
r_*\,\equiv\,\left (r_M\,r_c^2  \right )^{1/3}\,.
\label{int}
\eeq
This is a scale at which nonlinear interactions in a naive 
perturbative expansion in $G_N$ become comparable 
with the linear terms. 
 For a source such as the Sun, the hierarchy of the scales is given 
in (\ref {hierarchy}). Below, unless stated otherwise, we will consider 
sources  smaller than $r_*$. In Refs.~\cite{Andrei,MassimoBH,Lue,Sio,Nicolis}  approximate 
solutions for such sources were found in  different 
regions of (\ref {hierarchy}). The main properties of the 
solution can be summarized as follows:

\vspace{1mm}

(a) At distances $r\gg r_c$ the 5D Schwarzschild solution
with the 5D ADM mass $M$ is recovered (throughout this work $r$ stands for 
a 4D radius). 

(b) For $r_* \ll r \ll r_c$ the potential scales as in the 
4D Schwarzschild solution.  However, relativistic gravity is a 
tensor-scalar theory that contains the gravitationally 
coupled scalar mode (i.e. the tensorial structure is that of a
5D gravitational theory which contains extra polarizations). 

(c) For $r\ll r_*$  the theory reproduces 
the Schwarzschild solution of 4D general relativity (GR) 
with a good accuracy.

\vspace{1mm}

Perhaps the most important property of the (a-c) solution 
outlined above is the dynamical ``selfshielding'' 
mechanism by which the solution 
protects itself from the would-be strong coupling regime \cite {DDGV}.
Very briefly, the selfshielding can be described as follows: the 
expansion in $G_N$ breaks down at the scale 
$r\sim r_*$ making the perturbative calculations
unreliable below this scale. However, exact nonlinear solutions 
of equations of motion -- which effectively resum the series of classical 
nonlinear graphs -- are perfectly sensible well  below the scale $r_*$.
Hence, the correct way of doing the perturbative calculations is first 
to find a classical background solution  of equations of motion 
and then expand around it.

In Ref. \cite {GI}  a 4D part of the metric was exactly found. 
This exact result, 
combined with reasonable boundary conditions in the bulk, 
is sufficient to determine unambiguously a number of crucial properties 
of the solution.  First, this result confirms the existence of the 
scale $r_*$ -- this scale enters manifestly our exact solution. 
It also confirmed that the selfshielding mechanism outlined 
above takes  place. Furthermore, it was emphasized that 
the selfshielding effect takes 
place because a source creates a nonzero scalar curvature 
that extends {\it outside}  the source to a distance 
$r\sim r_*$.  This curvature suppresses nonlinear interactions 
that otherwise would become strong at the scale below $r_*$.
On the other hand, we also find that some of the physical properties 
of our solution differ from those in (a-c). The solution 
of Ref. \cite {GI}  has the 
following main features:

(A) For $r\gg r_c$, like in (a),  one recovers the 5D 
Schwarzschild solution, however unlike in (a), the 
new solution has the {\it screened} 5D ADM mass
\beq
M_{\rm eff}\, \sim \,M\, \left ({r_M\over r_c}\right )^{1/3}\,. 
\label{screenedmass}
\eeq
The screened mass is suppressed compared to the bare mass 
$M$. Therefore, the new  solution is  energetically favorable 
over the (a-c) solution.

(B) For $ r_*\ll r \ll r_c$ one can think of the solution 
as being a four-dimensional one with an $r$-dependent decreasing 
mass $M(r)\sim r_* r_M /r$. Alternatively, one can simply think of 
the solution just approaching very fast the 5D Schwarzschild metric with 
the screened mass (\ref {screenedmass}), i.e., approaching the 
asymptotic behavior of (A).

(C) For $r\ll r_*$ the results agree with those of (c) with 
a good accuracy. 

The (a-c) and (A-C) solutions both asymptote to the Minkowski space
at infinity. However, the way they approach the flat space is different
because of the difference in their 5D ADM masses. 
The (A-C) solution, or any of its parts,  cannot be obtained in the
linearized theory -- it is a nonperturbative solution at any distance scale.
Since the mass of the (a-c) solution is  larger than the mass 
of the (A-C) solution, we would expect that the heavier solution 
will eventually  decay into the light one, unless   
topological arguments prevent this decay.
 
The above findings suggest that the Minkowski space, 
although globally stable in the DGP model, is {\it locally} 
unstable in the following sense. A static source
placed on an empty brane creates a nonzero 
scalar curvature around it. For a source of the size $\lsim r_*$
this curvature extends to a distance $\sim r_*$. Above this scale 
the solution asymptotes very  quickly to a 5D Minkowski space. 
More intuitively, a static source distorts  a brane medium 
around it creating a potential well, and the distortion 
extends to a distance $r\sim r_*$. Since $r_*$ is much bigger than 
the size of the source itself, we can interpret this phenomenon as 
a local instability of the flat space. 
This local instability, however, has not been seen in the linearized 
theory \cite{DGP}. It should emerge, therefore, in nonlinear 
interactions and should disappear  when the scale $r_c^{-1}$ tends 
 to zero.\footnote{The latter assertion is valid 
since the (A-C) solution, as we will see,  
is regular in the $m_c\to 0$ limit where it turns into 
a conventional 4D solution, i.e. it 
has no vDVZ discontinuity \cite{Veltman,Zakharov}.}

It is remarkable that the distance scale to 
which the local instability extends, coincides with the scale
$r_*$ at which the naive perturbative expansion in $G_N$
breaks down.  Therefore, by creating a scalar 
curvature that extends to $r\sim r_*$,
the source shields itself from a would-be strong coupling 
regime that could otherwise appear at distances  
$r\lsim r_*$ \cite {DDGV}:
(i) The coupling of a phenomenologically 
dangerous  extra scalar polarization of a 5D graviton to 4D matter
gets suppressed at distances $r\lsim r_*$ due to the curvature
effects. This is similar to the suppression of the extra 
polarization of a massive graviton in  the AdS background 
\cite{KoganAdS,PorratiAdS}. Indeed, in our case the curvature 
created by the source, although coordinate dependent,  
has the definite sign that coincides with the sign of the 
AdS curvature. As a result, the model approximates 
with a high accuracy the Einstein gravity at $r\ll r_*$ with 
potentially observable small deviations 
\cite {Lunar1,Lue} (see comments below). 
(ii) The selfcoupling of the  extra polarizations 
of a graviton, which on a flat background leads to the breakdown of 
a perturbative expansion and to the strong coupling problem, gets now 
suppressed at distances $\lsim r_*$ by the scalar curvature created 
by the source. This is also similar to the suppression
of the selfcoupling of the massive graviton polarizations 
on the AdS background \cite {AGS,Luty}.

Above, we were primarily concerned with 
classical sources. Nevertheless, we would like to comment as well
on dynamics of ``quantum'' sources, such as gravitons.
Consider the following academic setup: 
a toy world in which there is no matter, radiation and/or any classical
sources of gravity -- only gravitons propagate and interact with 
each other in this world. Because of the very same trilinear 
vertex diagram that leads to the breakdown of the $G_N$ expansion for 
classical sources (see Ref. \cite {DDGV}), the selfinteractions of 
gravitons will become important at lower energy scale than they would 
in the Einstein theory. The corresponding breakdown scale is the 
scale (\ref {int}) adopted to  a quantum source with 
$r_M =1/\mpl$, that is $\Lambda_q^{-1}\sim (r_c^2/\mpl)^{1/3}$ \cite {Luty} 
(see also Ref. \cite {Rubakov} that obtains a somewhat 
different scale). In this setup the graviton loop diagrams
could in principle generate higher-derivative operators that are 
suppressed by the low scale. A theory with such  
high-derivative operators would not be predictive 
at distances below $\Lambda^{-1}_q\sim $ 1000 km or so. 

However, there are two sets  
of arguments suggesting that the above difficulty 
might well be unimportant for the description  of 
a real world which, on top of the gravitons, 
is inhabited by planets, stars, galaxies etc. 
We start with the arguments of Ref. 
\cite {Nicolis}. This work takes a point of view that 
$\Lambda_q$ is a true ultraviolet (UV) cutoff of the theory in a 
sense that at this scale some new quantum gravity degrees of freedom should
be introduced in the model. Nevertheless, as was discussed in 
detail in Ref. \cite {Nicolis},
this should not be dangerous if  one considers a realistic 
setup in which mater  is  introduced into the theory. For instance, 
consider the effect of introducing the classical gravitational field 
of the Sun. Because of the gravitational background of the Sun, 
the UV cutoff of the theory becomes a coordinate dependent
quantity $\Lambda_q(x)$. This cutoff grows closer to the source 
where its gravitational field  becomes more and more pronounced,
hence, increasing the value of the effective UV cutoff.
In this approach the authors of Ref. \cite {Nicolis} managed 
to find a minimal required set  of higher-dimensional operators that 
are closed with respect to the renormalization group  flow. 
Because of the resummation of  large classical nonlinear effects 
these operators are effectively suppressed by the 
coordinate-dependent scale $\Lambda_q(x)$.  If so, the new UV physics will not
manifest itself in any measurements \cite {Nicolis}.

Putting all this on a bit more general ground, 
one should {\it define} the model in an external 
background field.  That is, in the action and the partition function 
of the model the metric splits into  two parts $g_{\mu\nu}=
g^{\rm cl}_{\mu\nu} +g^{\rm q}_{\mu\nu}$, 
where $g^{\rm cl}_{\mu\nu}$ stands for the classical background 
metric and $g^{\rm q}_{\mu\nu}$ denotes the quantum fluctuations 
about that metric. The classical part  
satisfies the classical equations of motion with given classical 
gravitational sources such as planets, stars, galaxies etc..
Then, the effective UV cutoff for quantum fluctuations 
at any given point in space-time 
is a function of the background metric. For a 
realistic setup this effective cutoff is high enough to 
render  the model consistent with observations.

We find the above logic useful and viable. We also think that
the algorithm of Ref. \cite {Nicolis} might be the most convenient 
one for practical  calculations. Nevertheless, there could 
exist deeper dynamical phenomena  beyond 
the above approach to the discussions of which we turn right now. 
Although our arguments below parallel in a certain  respect 
those of Ref. \cite {Nicolis}, there is a conceptual difference on 
the main issue. Our view, that we will try to substantiate in subsequent 
works, is that the scale $\Lambda_q$ is not a UV scale of the 
model in the sense that some new  quantum gravity degrees of freedom 
should be entering at that scale. We think that all what's needed 
to go above the scale  $\Lambda_q$ is already in the model,
and that this is just a matter of 
technical difficulty of nonperturbative calculations
(or, in other words, is a matter of difficulty of summing up
loop diagrams). The resummation could in principle cure 
problems at the loop level as well. At this end, we do not see a reason why  
the selfshielding mechanism outlined above  should not be 
operative for ``quantum'' sources too. 
The very same local instability of the Minkowski space 
should manifest itself in nonlinear interactions of quantum 
sources, e.g., gravitons.  The 
local instability scale in that case is $\Lambda_q$. Hence, we 
would expect that a quantum source creates a curvature around it 
that extends to the distances of the order of $\sim (r_c^2/\mpl)^{1/3}
\sim 1000$ km, and doing so it selfshields itself from the strong 
coupling regime. If this is so, then the problem of loop calculations
boils down to the problem of defining correct variables with respect to which 
the perturbative expansion should be performed. In this case the 
field decomposition should take the form:
$g_{\mu\nu}=g^{\rm np}_{\mu\nu} +g^{\rm q}_{\mu\nu}$, 
where $g^{\rm np}_{\mu\nu}$ stands for a nonperturbative 
background metric created by a ``quantum'' source.  
Similar in spirit arguments using a toy model  
were given by Dvali in Ref. \cite {Dvali}.

We find it  useful to adopt a gauge in which the 
line element has an off-diagonal form:
\beq
ds^2= {\rm e}^{-\lambda} dt^2-{\rm e}^{\lambda} dr^2\, -
\,r^2d\Omega^2-\gamma\, 
drdy -{\rm e}^\sigma dy^2~,
\label{lineelement}
\eeq
where $\lambda,\  \gamma,\ \sigma$ are functions of $r$ and $y$.
Our brane is located at $y=0$ in this coordinate system. 
The ${\bf Z}_2$ symmetry w.r.t the brane 
implies that $\gamma $ is an odd function of $y$ while the 
rest are even.  A more conventional diagonal coordinate system 
can be obtained  by a coordinate redefinition after which the interval reads
\begin{equation}
ds^2={\rm e}^\nu dt^2-{\rm e}^\lambda dr^2- r^2d\Omega^2-{\rm e}^
{\beta} dz^2~.
\label{common}
\end{equation}    
Here the functions $\nu $ and $\beta$ are related 
to  $\lambda, \gamma,$ and $ \sigma $. In the $z,r$ coordinate system 
the brane is bent. Typically in the brane-world models the 4D part of the Einstein 
equations are not closed. Hence, the induced metric on a brane 
cannot be determined without some input from the bulk equations,
and/or without making certain assumptions about the induced metric itself.
This would also be true in our case. However, in the gauge 
(\ref {lineelement}), 
we find a subset of the Einstein equations that 
can be closed for the function $\lambda$.  As a result, $\lambda$  can be 
found exactly on the brane. Although the knowledge of $\lambda$ alone is not 
enough to describe all gravitational dynamics on the brane (
for instance, this is not enough  for the description  of 
the matter geodesics at short distances since transverse derivatives 
of the metric are also entering the 5D geodesic equations) nevertheless, 
combining the knowledge of $\lambda$ with  the 
asymptotic behavior of the other functions in (\ref {lineelement}) 
that we can also obtain unambiguously, is enough to deduce the 
properties (A-C). Hence, these properties are ``exact.''

Finally, we would like to make  two important comments.
First, the DGP model possesses two branches of solutions 
that are distinguished from each other  
by the bulk boundary conditions. These two branches are 
disconnected.
In this work we concentrate primarily on  the Schwarzschild solution of 
the so called conventional branch on which the brane and the bulk asymptote 
to the Minkowski space at infinity.
However, the second, the so called ``selfaccelerated'' branch 
\cite {Cedric} 
is extremely interesting as it can be used to describe the 
accelerated expansion of the Universe without introducing dark 
energy \cite {DDG}. In the present work we also find an
exact brane metric for a Schwarzschild source  
on the selfaccelerated branch. However, because the asymptotic 
behavior of the solution on this branch is not Minkowski we are not 
able to argue for the existence of a nonsingular bulk solution. 
On the other hand, we do not see any physical reason why this solution 
should not exist in the bulk as well. This branch will be discussed
in detail elsewhere.  Second, it is interesting to note that 
the linearized analysis 
of the DGP model in dimensions six and higher \cite {GGMisha}, 
as well as certain modifications of the five-dimensional 
model \cite {Romb,GG} show no sign of breakdown of perturbation theory
and strong nonlinear effects. It is left for future work to understand
more deeply the interconnections between all these approaches. 

\subsubsection{Structure of the solution}

In this subsection we discuss the properties of the solutions 
on the brane, i.e., 
at $y=0$. In this discussion we closely follow Ref. \cite {GI}.
We find  certain similarities, as well as  drastic 
differences, in the 4D part of our solution with  the 
anti-de Sitter--Schwarzschild 
(AdSS) solution of conventional 4D 
General Relativity  (GR) with  a small positive 
cosmological constant $\Lambda$ (this is in spite of the fact that 
a source in our solution creates a curvature that 
has a signature of a negative cosmological constant) 
\beq
G_{\mu\nu}\,+\,\Lambda g_{\mu\nu}\,=-\,8\,\pi\, G_N\,T_{\mu\nu}\,.
\label{EL}
\eeq
It is instructive to contrast our solution to the AdSS metric.

Consider 4D GR with  the cosmological constant $\Lambda =-3m_c^2$. 
Furthermore, 
consider a static source of mass $M$ (a star) and a Schwarzschild radius 
$r_M\equiv  2G_N M$ in this space. In the static coordinate system 
the AdSS solution takes the form
\beq
ds^2 &= &\left (1\,-\,{r_M\over r}\,+\,m_c^2\,r^2\right )\,dt^2
\nonumber\\[3mm]
 &-&
{dr^2 \over \left (1\,-\,{r_M\over r}\,+\,m_c^2\,r^2\right )  }\,-
\,r^2\,d\Omega_2^2\,.
\label{dSS0}
\eeq 
This coordinate system covers  the AdSS solution  in the interval 
$$r_M < r< r_c \equiv m_c^{-1}\,.$$  The following  properties of the AdSS 
solution will be contrasted to our solution.

(i) In the interval $r_M<r<r_c$ there is a new distance scale $r_*$
(\ref {rstar}) exhibited by (\ref {dSS0}).
The physical meaning of this scale is as follows.
For $r<r_*$ the Newtonian potential $r_M/r$ in (\ref {dSS0}) 
dominates over the term  $m_c^2r^2$,  while for $r>r_*$ the 
term   $m_c^2r^2$ overcomes  the Newtonian term. Hence,  
$r_* $ is a scale at which the Newtonian and the $m_c^2r^2$ terms 
are equal. This can  also be expressed in terms of invariants. 
Let us define the Kretschman  scalar (KS)
\beq
R_K\,\equiv\,\sqrt{(R_{\alpha\beta\gamma\delta}^{Sch})^2}\,,
\label{RK}
\eeq 
where $R_{\alpha\beta\gamma\delta}^{Sch}$ is a Riemann tensor of the 
Schwarzschild part of the solution (i.e., of the part that 
survives in the $m_c \to 0$ limit). 
We compare the KS  with the background curvature due to 
the cosmological constant
\beq
|R_{\Lambda}|\,=\,12\,m_c^2\,.
\label{rmc}
\eeq 
We get
\beq
R_K\,\gg |R_{\Lambda}|\, ~~~{\rm for}~~r\ll r_*\,; ~~~~
R_K\,\ll |R_{\Lambda}|\, ~~~{\rm for}~~r_*\ll r \ll r_c \,.
\label{curvatures}
\eeq 
Therefore, $r_*$ is a scale at which $R_K \simeq  |R_{\Lambda}| $.
For $r_M \ll r \ll r_*$ the corrections due to the background curvature 
are small and the solution is dominated by the Schwarzschild metric, 
while for $r_*\ll r\ll r_c$ the background curvature terms are larger that 
the Schwarzschild  terms, both of them still being smaller than 1.

(ii) At $r>r_c$ the Schwarzschild part becomes irrelevant 
compared to the AdSS part. 

We will show below that our solution has some of the 
properties described in (i), however, unlike (ii), it 
behaves as 5D Schwarzschild solution at large distances.

The 4D part of our solution (i.e. the solution at $y=0$) 
for $r\ll r_c$ takes the form
\beq
ds^2|_{y=0} &=& \left (1- {r_M\over r} + m_c^2 r^2 g(r)\right )dt^2 
\nonumber\\[3mm]
 &-& {dr^2 \over \left (1 -{r_M\over r} + m_c^2 r^2 g(r)\right )  }-
\,r^2\,d\Omega_2^2\, .
\label{sol4D}
\eeq 
Like the AdSS solution, the metric (\ref {sol4D})  possesses the $r_*$ 
scale defined in (\ref {rstar}). As we will see below this scale has the 
same physical meaning as in the AdSS case.  For instance,  at $r\ll r_*$ 
\beq
g(r) \,\simeq  \,\left 
({r_*^4\over r^4}\right )^{1\over 1+\sqrt{3}}\,. 
\label{grstar}
\eeq 
Then, it is straightforward to check that 
\beq
{r_M\over r} \gg m_c^2\,r^2 g(r)\,~~~ {\rm for } ~~~ r\ll r_*;~~~~~
{r_M\over r}\, \sim \, m_c^2\,r^2 g(r)\,~~~ {\rm for } ~~~ r\sim r_*\,.
\label{curvsol}
\eeq 
Therefore, the corrections become of the order of the 
${r_M/r}$ term at around $r\sim r_*$.
Moreover, like the AdSS solution, the corrections 
dominate over  ${r_M/r}$ for $r_*\ll r \ll r_c$
turning the 4D behavior of the solution into the 5D 
behavior. The plot of the function is given on Fig. 3.
\begin{figure}
\centerline{\epsfbox{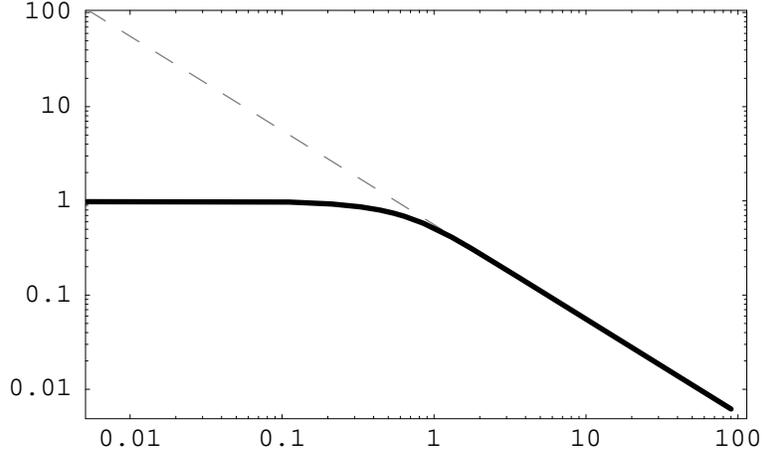}}
\epsfysize=6cm
\vspace{0.1in}
\caption{\small 
The solid line depicts $P(r)/r_M$ (on the 
vertical  axis, where $P\equiv r(1-g_00)$), 
as a function of $r$ on the horizontal axis. The dashed 
line presents  the function $0.28{r_*}/r$;  The value of 
$r_*$ is set  to 1 on this graph.}
\label{fig1}
\end{figure} 
As in the AdSS case, the corrections to the Schwarzschild solution 
that are proportional to $m_c$ give rise to the four-dimensional Ricci 
curvature $R_{m_c}$. This is interesting since the curvature is completely 
due to the modification of gravity. However, unlike the AdSS case, this 
curvature is not a constant but depends on $r$;  moreover it 
also depends on the strength of the source itself.
The plot of the Ricci curvature is given on Fig. 4.
\begin{figure}
\centerline{\epsfbox{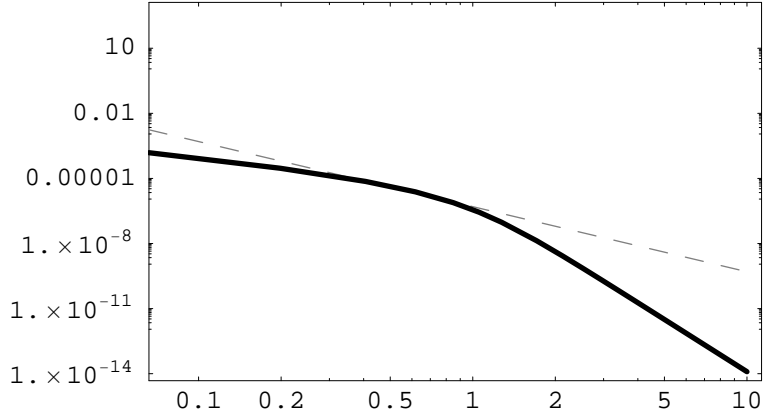}}
\epsfysize=6cm
\vspace{0.1in}
\caption{\small 
The solid line depicts the magnitude of the 
four-dimensional Ricci scalar curvature
(on the vertical axes) as a function of $r$ on the horizontal axes.
The dashed line depicts the dependence of the 4D Kretschmann scalar
on $r$. The value of $r_*$ is set to 1 on this plot.}
\label{fig2}
\end{figure} 
The presence of this curvature can easily be understood
by looking at the trace of the 4D Einstein equation on the brane
\beq
R_4\,+\,3\,m_c\,K\,=\, 8\pi\,G_N T\,.
\label{traceK}
\eeq 
$T$ is zero outside a localized source such as a star. However, the 
trace of the extrinsic curvature $K$ is not zero, therefore, 
$R_{m_c}=-3m_c\,K\ne 0$ and $R_4$ outside of the source is nonzero and 
equals to $R_{m_c}$. 

Similar to the AdSS solution the above properties can be expressed 
in terms of the invariants
\beq
R_K\,\gg R_{m_c}~~{\rm for}~r\ll r_* ~~{\rm and}~r\gg r_*\,;
~~ R_K\,\sim  R_{m_c}~~{\rm for}~r\sim r_*\,.
\label{compcurv}
\eeq 
Unlike the AdSS solution, however,  
the curvature $R_{m_c}(r)$  decreases very fast 
after $r\gg r_*$. Hence, the induced curvature $R_{m_c}(r)$ is subdominant
to $R_K$ everywhere except in the neighborhood of the point $r\sim r_*$
where they both are of the same order $\sim m_c^2$, see Fig. 4.

Furthermore, unlike the AdSS solution, our 
solution can be presented in the same coordinate system even for $r\ge r_c$. 
This is because there is no horizon at $r=r_c$ and our solution smoothly 
turns into the 5D Schwarzschild solution
\beq
ds^2|_{y=0} \,=\, \left (1\,-\,{{\tilde r}^2_M\over r^2}\right )\,dt^2 \,
-\, {dr^2 \over \left (1\,-\, {{\tilde r}^2_M\over r^2} \right )  }\,-
\,r^2\,d\Omega_2^2\,.
\label{sol5d}
\eeq 
The key  property of this solution is that the gravitational radius is rescaled
\beq
{\tilde r}_M \,\sim\, r_M\,\left ({r_c\over r_M}     
\right )^{1/3} \,\gg\,r_M\,.
\label{5DrM}
\eeq 
This has an explanation. The gravitational radius grows compared to $r_M$ 
because in the 5D regime the gravitational coupling constant grows. 
However, there is an opposite effect as well. 
In fact, the gravitational radius 
reduces compared to what it would have been in a pure 5D theory with no brane.
This is because the effective mass of the source $M_{\rm eff}$, defined as 
${\tilde r}_M^2 \equiv M_{\rm eff}/M_*^3$, gets screened.  Indeed,  
$M_{\rm eff}$ includes contributions from the curvature $R_{m_c}(r)$ 
that stretches out all  the way to $r=r_c$.  The screened mass of the 
source is
\beq
M_{\rm eff} \,\sim \,M\,\left ({r_M\over r_c}     \right )^{1/3} \,\ll \,M\,.
\label{Meff}
\eeq 
Thus, as seen from $r\gg r_c$ distances, there is strong screening of the 
source. 

All the above results could be understood as follows. Consider an empty brane
and an empty bulk. The Minkowski space is a solution. Let us place a 
static source on the brane at $r=0$. The Minkowski solution remains 
globally stable, however, the source, no matter how weak, triggers local 
instability of the Minkowski space on a brane in the region $r\le r_*$.
In this patch  the Minkowski space is readjusted to a curved space. The 
curvature of the latter depends on the strength of the source, it   
slowly decreases with increasing $r$ but drops fast at  $r> r_*$. 
For an observer at large distances  it looks as if  the  source has  
polarized the medium (brane) around it. This observer measures the 
screened mass (\ref {5DrM}) which also includes the contributions 
of the curvature.

At large enough distances, i.e. $\sqrt{r^2+y^2}\gg r_*$, 
the solution turns into a 5D spherically-symmetric 
Schwarzschild solution,
\beq
ds^2|_{\sqrt{r^2+y^2}\gg r_*} 
&\sim& \left (1\,-\,{{\tilde r}^2_M\over r^2\,+\,y^2}\right )\,dt^2 \,
\nonumber\\[3mm]
&- & {dr^2 \over \left (1\,-\, {{\tilde r}^2_M\over r^2\,+\,y^2} \right )  }
\,- \,(r^2+y^2)\,d\Omega_3^2\,.
\label{bulk5D}
\eeq
However, the 5D spherical symmetric is only an approximation
and does not hold for  $\sqrt{r^2+y^2}\ll r_*$. 
In the latter regime the properties of the solution 
on and off the brane are rather different. The pure 5D spherically-symmetric 
solution (\ref {bulk5D}) 
is squeezed both  on and off the brane but it is more squeezed
on the brane than in the bulk. Hence, the only symmetry of the 
solution that is left is the cylindrical symmetry.

\section{Brane-induced gravity in more than five  
dimensions -- softly massive gravity}

In the present section we turn to Brane Induced Gravity in
more than 5 dimensions. These models share the main property of the 
5D model that gravity at short distances is four-dimensional, 
becoming higher dimensional at  larger scales. However, there are a 
number of crucial distinctions from the 5D case. These are:

\begin{itemize}

\item{Unlike a 3-brane embedded in 5D space-time, 
the embedding into six and higher dimensional space-times allows
to preserve the 3-brane world-volume to be flat. This suggest that the 
CCP could be solved in this approach in dimensions six and higher
\cite {DG,DGS}.}

\item{Unlike the 5D case, the transverse to the brane 
Green's functions in six dimensions and higher are sensitive to the 
UV physics. A careful treatment  of  the UV regularization procedure
\cite{DG,Wagner,Kiritsis,DGHS} is needed. The best option is to use
the UV completion dictated by  the String Theory construction \cite {Ignatios}.}

\item{In most of the interesting cases where  simplest 
UV regularizations were tried so far, the flat space propagator 
exhibits new poles that correspond to very light $m \sim H_0$ 
tachyons \cite {Kaku} with negative residues \cite {Siopsis}. 
The positions of these  poles and their mere existence is UV 
regularization dependent. At present, it is not clear whether these poles 
would persist in a consistent UV completed  theory, such as in the 
string theory construction \cite {Ignatios}. Work to establish this 
is in progress \cite {Ignatios2}. However, even in the regularizations 
where  the new poles exist, they can be eliminated by a special 
prescription for the poles of the Greens functions \cite{GGMisha} 
at the expense
of sacrificing causality at the Hubble scales. This suggest that 
if these poles are truly present  in the theory, 
then the flat space-time, although globally stable,
might not be a locally stable background within a patch of the size 
$H_0^{-1}$. Such a local instability takes place in the 5D model 
but only at a nonlinear level and at a different scale \cite {GI}.
This issue in the context of six- and higher-dimensional models 
awaits further investigation.}

\item{Unlike the 5D case, it turns out that in six and higher dimensions 
perturbation theory does not break down at low scale \cite{GGMisha}. This 
observation is independent of the UV regularization as well as of the 
presence/absence of the new poles discussed in the previous item 
(i.e., the naive perturbation theory does not break down at a low scale even 
when the new poles are not present, see below)}.

\end{itemize}

The first issue on the list above was already discussed in Sect. 2. 
In what follows we will study in turn the second, third and fourth issues. 
As was mentioned above, these three 
items apply to a brane that has no tension (i.e, 
to a brane the cosmological constant on which is fine-tuned to zero).
However, our main goal is to study a brane with an arbitrary 
tension, and the last three properties listed above 
should be studied and established for  
this case. This is  only partially fulfilled so far,  
with certain encouraging results (see below). Further detailed 
calculations are still needed. Here we derive results for a 
tensionless brane following Ref. \cite {GGMisha}.

The equation of motion for the theory described by the action 
(\ref {1}) in $D\ge 6$ takes the form 
\beq
\d \, \mpl^2 \, G^{(4)}_{\mu\nu}\,\delta^\mu_A\,\delta^\nu_B
\,+\, \m^{2+N} \,G^{(D)}_{AB}\,
= \, - T_{\mu\nu}(x)\, \delta^\mu_A\,\delta^\nu_B \, \d \,.
\label{geq0}
\eeq
$G^{(4)}_{\mu\nu}$ and $ G^{(D)}_{AB}$ 
denote the four-dimensional  and
$D$-dimensional Einstein tensors, respectively. 
We choose (for simplicity)  a source  localized on 
the brane, $T_{\mu\nu}(x)\d $.

Gravitational dynamics encoded in Eq.~(\ref {geq0})
can be inferred  both  from the four-dimensional (4D)
as well as $(4+N)$-dimensional standpoints.
 From the 4D perspective,  gravity on the brane is mediated
by an infinite number of the Kaluza--Klein  (KK) modes that
have no mass gap. Under  conventional circumstances
(i.e.,  with no  brane kinetic term) this would
lead to higher-dimensional interactions. However,
the large 4D Einstein--Hilbert (EH) term
suppresses the wave functions of  heavier KK modes,
so that in effect they  do not participate  in the
gravitational interactions  on the brane at observable distances
\cite {DGKN}.  Only  light KK modes,
with masses $m_{KK}\lsim m_c$,
\beq
m_c \equiv \frac{\m^2}{\mpl}\,,
\label{mc}
\eeq
remain essential,
and they collectively act  as an effective 4D graviton with
a typical mass of the order of $m_c$ and a certain smaller width.

Assuming that $\m\sim 10^{-3}$ eV
or so, we obtain  $m_c \sim H_0\sim 10^{-42}$ GeV.
Therefore, the DGP model with $N\geq 2$ 
predicts~\cite {DGHS} a modification of gravity 
at short distances $\m^{-1}\sim 0.1$ mm {\em and} at large 
distances  $m_c^{-1}\sim
H_0^{-1}\sim 10^{28}\,\,{\rm cm}$, give or take an order of magnitude.
Since gravitational interactions,  nevertheless, are mediated
by an infinite number of states at  arbitrarily low
energy scale,  the effective theory  (\ref {1}) presents, from the 4D 
standpoint, a {\it nonlocal} theory~\cite {DGS}. Moreover,
as was suggested in \cite {ADDG}, nonlocalities postulated in 
{\em pure  4D} theory  can solve the CCP \cite {ADDG}, 
and give rise to  new mechanisms  for the present-day acceleration 
of the universe~\cite{ADDG}. (It is interesting to note that the nonlocalities 
in a gravitational theory that are needed to solve the 
cosmological constant problem 
could appear from quantum gravity \cite {Veneziano} or matter 
loops \cite {Mottola} in a purely 4D context.)

On the other hand, from the $(4+N)$-dimensional perspective,
gravitational interactions are  mediated by a single
higher-dimensional graviton. This graviton has two kinetic terms
given in Eq.~(\ref {1}), and, therefore, can propagate differently
on and off the brane. Namely, at short distances, i.e. at
$r< m_c^{-1} \sim H_0^{-1}\sim 10^{28}\,\,{\rm cm}$,
the graviton emitted along the brane
essentially propagates along  the brane and mediates
4D interactions. However, at larger distances, the
extra-dimensional effects take over and gravity
becomes ($4+N$)-dimensional.

As was first argued in Ref. \cite {DG}, the results in 
$N\ge 2$ DGP models are sensitive to  ultraviolet (UV) physics, 
in contradistinction to  the $N=1$ model \cite {DGP}. 
In other words, one should either consistently smooth out the 
width of the brane \cite {DGHS}, or introduce a 
manifest UV cutoff in the theory \cite {Wagner,Kiritsis,DGHS}, 
or do both. With a finite thickness,  more localized 
operators  appear on the world-volume of the brane, 
in addition to  the world-volume Einstein--Hilbert term 
already present in  Eq.~(\ref {1}) \cite {DGHS}. For instance, one could  
think of a higher-dimensional Ricci scalar 
smoothly spread over  the world-volume \cite {Romb}.

In general, terms that are square of the extrinsic curvature can 
also emerge. Some of these
terms can survive in the limit when the brane thickness 
tends to zero (i.e. in the low-energy approximation).
For instance, in the zero-thickness limit  of the brane
the following terms might  be important:
\beq
\d\, h^\mu_\mu \,\partial_\alpha^2 \, h^a_a\,,~~~
\d\, h^{\mu\nu}\,\partial_\mu \partial_\nu \, h^a_a\,,~~~\d \,h^\mu_\mu
\, \partial_a \partial_b \, h^{a b}\,,
\label{terms}
\eeq
where $h$ denotes small perturbations on flat space.
Although the main features of the model, such as  
interpolation between the 4D power-law behavior of a nonrelativistic 
potential at short distances and the 
higher-dimensional behavior at large distances,
are not expected to be changed by adding these terms, 
nevertheless, the tensorial structure of a propagator 
could in general depend on these terms and 
selfconsistency of the theory may 
require some of these terms to be present in the actions in a 
reparametrization invariant way.

In the low-energy approximation the exact form of these ``extra" terms 
and their coefficients  are ambiguous, because of their UV origin.
They will be fixed in a fundamental theory 
from which the DGP model can be derived \cite {Ignatios,Ignatios2}.
In the present paper, in the absence of such a fundamental theory, 
(but in the anticipation of its advent),
we would like to study a particular parametrization of these ``extra" 
terms,
for demonstrational purposes. According to our
expectations, physics in the 
selfconsistent theory will have   properties very similar 
to those discussed below. We will show that these properties are 
rather attractive since they do avoid severe problems of 4D 
massive gravity. 
 
Consider the action 
\beq
S&=&{\mpl^2 } \,\int\,d^4x\,\sqrt{g}\,\left ( a \,R(g)\,+\,
b \, {\mathcal R}_{4+N} \right )
\nonumber\\[3mm]
&+&
{\m^{2+N}}\,\int \,d^4x\,d^Ny\,\sqrt{{\bar g}}
\,{\mathcal R}_{4+N}({\bar g})\,,
\label{11}
\eeq
where, in addition to the 4D EH term,   a 
D-dimensional EH term localized on the brane is present. Here $a$ and $b$ are 
some numerical coefficients. 
We will study the properties of the system described by (\ref{11})
for different values of $a$ and $b$. The action (\ref{11})
is fully consistent with the philosophy of Ref.~\cite{DGP}:
if there is a (1+3)-dimensional brane in D-dimensional space,
with some ``matter fields" confined to this brane,
quantum loops of the confined matter will induce
all possible structures consistent with the
geometry of the problem, i.e. (1+3)-dimensional wall
embedded  in D-dimensional space.

The equation of motion in the  model (\ref {11}) takes 
the form 
\beq
&&\d \, \mpl^2 \, \left ( a\,G^{(4)}_{\mu\nu}\,+\, b\,G^{(D)}_{\mu\nu} \right )
\delta^\mu_A\,\delta^\nu_B
\nonumber\\[3mm]
&&+\, \m^{2+N} \,G^{(D)}_{AB}\,
= \, - T_{\mu\nu}(x)\, \delta^\mu_A\,\delta^\nu_B \, \d \,.
\label{geq}
\eeq
In deriving the above equation we first 
introduced a finite brane width $\Delta$, 
and then took the $\Delta \to 0$ limit in such a way 
that  no surface terms appear. In general, the results 
depend on the regularization procedure for the brane width. 
In the present work we adopt a simple prescription in which 
derivatives with respect to the transverse coordinates calculated 
on the brane vanish in the $\Delta \to 0$ limit (a unique prescription 
could only be specified by a fundamental theory.).
As previously,  $G^{(4)}$ and $ G^{(D)}$ 
denote the four-dimensional  and
$D$-dimensional Einstein tensors, respectively, while
  $a$ and $b$ are certain constants.
In order to be able to describe 4D gravity 
at short distances with the right value of 
Newton's coupling we set
\beq
a\,+\,b\,=\,1\,.
\label{ab}
\eeq
Note, that the first two terms in parenthesis 
on the left-hand side of Eq.~(\ref {geq}) can be identically 
rewritten as 
\beq
&& (a+b) \, G^{(4)}_{\mu\nu}\,+\, 
b\left (-\partial_\mu \partial_a \, h^a_\nu \,- \,
\partial_\nu \partial_a \, h^a_\mu \,- \, \partial_a^2 h_{\mu\nu}\, + \,
\eta_{\mu\nu}\partial_a^2 h^C_C  \right.\nonumber\\[2mm]
&&+ \left.\partial_\mu \partial_\nu 
h^a_a \, -\,\eta_{\mu\nu} \partial_\alpha^2\,h^a_a\, + \, 2\eta_{\mu\nu}
\partial_a\partial_\alpha h^{a\alpha}\,+\,\eta_{\mu\nu}\partial_a\partial_b
h^{ab}\right) .
\eeq
The above  equation of motion (\ref {geq}) --
which should  be viewed as a regularized version of the DGP model --
could  be  obtained from the action   (\ref {1}) as well, provided
the latter is
amended by certain extrinsic curvature terms. Below we will study this 
version of the regularized DGP model for  various values  
of the parameters $a$ and $b$.  

\subsection{Perturbation theory of flat space}

We start our discussion with a simple model of
a scalar field $\Phi$ in $(4+N)$-dimensional space-time.
For convenience we separate
the dependence of the scalar field $\Phi$ on four-dimensional and
higher-dimensional coordinates,
$\Phi (x_\mu, y_a) \equiv \Phi(x,y)$.
The  two kinetic term action  -- the scalar counterpart of
Eq.~(\ref {1})  -- is
\beq
S&=& {\mpl^2}\, \int d^4x\,\partial_\mu\Phi(x,0)\,
\partial^\mu\Phi(x,0)
\nonumber\\[3mm]
&+& {\m^{2+N}} \, \int d^4x\,d^Ny\,
\partial_A\Phi(x,y)\,\partial^A\Phi(x,y)\,.
\label{saction}
\eeq
It is important to understand that in
the scalar case the analog of the new term  
included in Eq.~(\ref {geq}) but absent in  (\ref {geq0})
reduces, identically, to the already  existing localized term.
This is a consequence of  our choice of the  regularization of 
the brane width $\Delta$ and  the boundary conditions
according to which  transverse derivatives vanish on the brane in the 
$\Delta \to 0$ limit. 

To study interactions mediated by the scalar field
we assume that $\Phi$  couples to a source $J$ localized in the
4D subspace in a conventional way, $\int d^4x 2 \Phi(x,0)\,  J(x)$.
Then the  equation  of motion takes the form
\beq
\d \, \mpl^2 \, \partial^2_\mu \,\Phi(x,0)\,+\,
\m^{2+N} \,\partial^2_A \, \Phi(x,y)\, = \, J(x)\, \d \,.
\label{seq}
\eeq
The very same equation applies to the scalar field Green's
function.

To  solve this equation it is convenient to  Fourier-transform
it  with respect to ``our" four space-time coordinates
$x_\mu\to p_\mu$, keeping the extra $y$ coordinates intact. Marking 
the Fourier-transformed quantities by the tilde, 
\beq
\Phi(x,y)\, \to \, {\tilde \Phi}(p,y)\,,
\label{fourier}
\eeq
we then get from Eq.~(\ref{seq})
\beq
&&\d \, \mpl^2 (-p^2)\,{\tilde \Phi}(p,0)
\nonumber\\[2mm]
&&+\,
\m^{2+N} \,(-p^2 -  \Delta_y ) {\tilde \Phi}(p,y)\, =
\,{\tilde J}(p)\, \d \,,
\label{seqmom}
\eeq
where $p^2\equiv p_0^2-p_1^2-p_2^2-p_3^2$, and  the notation
\beq
\Delta_y \, \equiv \, \sum_{a=1}^N {\partial^2\over \partial y^2_a}
\label{Delta}
\eeq
is used.

We will look for the solution of Eq.~(\ref {seqmom}) in the 
following form:
\beq
{\tilde \Phi}(p,y)\,\equiv \, D(p,y)\,\chi(p) \,,
\label{decompos}
\eeq
where the function $D$ is defined as a solution of the equation
\beq
(- p^2 -  \Delta_y -i{\bar \e})\, D(p,y)\,= \, \d\,.
\label{D}
\eeq
Note that the function $D$ is uniquely determined only after 
 the $i{\bar \e}$ prescription  specified above is implemented.
We also introduce a convenient abbreviation
\beq
D_0(p)\, \equiv \,D(p, y=0)\,.
\label{D0}
\eeq

Now, it is quite obvious that a formal solution of Eq.~(\ref {seqmom}) 
can  be written 
in terms of the function $D$ as follows:
\beq
{\tilde \Phi}(p,y) \,=\,-\,{{\tilde J}(p)  \over \mpl^2}\,\,\,
{D(p, y) \over p^2\,D_0(p)\, - \,
\m^{2+N}/\mpl^2 }\,+c\,{\tilde \Phi}_{\rm hom}(p,y)\,,
\label{ssol}
\eeq
where ${\tilde \Phi}_{\rm hom}(p,y)$ is a general
solution of the corresponding  homogeneous equation
(i.e., Eq. (\ref {seqmom}) with the vanishing right-hand side),
and $c$ is an arbitrary constant. Equation (\ref{ssol})
presents, in fact, the Green's function too, up to 
the factor $\tilde J(p)/ \mpl^2$, which must be
amputated. In particular, for the Green's function
on the brane we have
\beq
G(p,0)=\frac{\mpl^2}{\tilde J(p)}\,   {\tilde \Phi}(p,y=0) \,,
\quad  G_{\rm hom}(p,0) =\frac{\mpl^2}{\tilde J(p)}\,
{\tilde \Phi}_{\rm hom}(p,y=0)\,,
\label{G}
\eeq
while for arbitrary values of $y$
\beq
G(p,y)\, = \,- \,{ D(p,y) \over p^2\, D_0(p) -u^N}\,
+\,c\,G_{\rm hom}(p,y)\, ,
\label{spoles}
\eeq
where
\beq
u^N \equiv \, \frac{\m^{2+N}}{\mpl^2}= m_c^2\, M_*^{N-2}\,.
\label{defu}
\eeq
The presence/absence of the homogeneous part is regulated by the $i\epsilon$
prescription.
Note that if the first term on the right-hand 
side of Eq.~(\ref {spoles})
has poles on the real axis of $p^2$, then the 
homogeneous equation has a solution
\beq
G_{\rm hom}(p,y)\,=\, D(p,y)\,\delta \left (p^2D_0(p) - 
u^N  \right )\,.
\label{homsc}
\eeq
This fact will play an essential role for gravity.

In what follows we will examine
the poles of  the Green's function $G(p,y)$.
The positions of these poles depend on
the functions $G_{\rm hom}(p,y)$, and
$D_0$ as defined in Eqs.~(\ref{D}) and  (\ref {D0}).
The choice of a particular rule of treatment of
the poles corresponds to the choice of appropriate 
boundary conditions in the coordinate space. Note that the latter are
dictated by physical constraints on the Green's function $G$ 
rather than on the auxiliary function $D$.

To get to the main point, we will try the 
simplest strategy of specifying the poles and check, 
{\em a posteriori}, whether  this strategy
is selfconsistent. 
Let us put
$$c=0$$
 and define $D$ in the Euclidean momentum space.
Since in the Euclidean space the expression for $D$ is well-defined
and has no singularities,
\beq
D(p_E, q) =\frac{1}{p_E^2 +q^2}\,,\qquad 
D(p_E, q)\equiv \int d^N y \, e^{iqy}\, D(p_E, y)\,,
\label{deuclid}
\eeq
$$
q^2 = \sum_a (q^a )^2\,,
$$
one can perform  analytic continuation from the Euclidean space 
to Minkowski.  This is not the end of the story, however.
It is the Green's function $G$ that we are interested in, not the
auxiliary function $D$. As  will be explained  below,
the above  procedure is consistent, for
the following reason.  The function $G$ obtained in this way
has a cut extending from zero to infinity. 
In addition, we find two complex conjugate 
poles on the second {\em unphysical} Riemann sheet of the complex 
$p^2$ plane. Moreover, there are additional poles on subsequent unphysical 
sheets. 

\subsection{Six dimensions}
\label{sd}

It is instructive to demonstrate how things work by
considering  separately the six-dimensional case.
In six dimensions  sensitivity to the UV cutoff is only
logarithmic, and it is conceivable  that the results obtained in the
cut-off theory could be consistently matched  to those
of a more fundamental UV-completed theory-to-come.\footnote{
The $D> 6$ models of brane-induced gravity 
are power sensitive to  UV physics. In general one expects 
all sorts of higher-derivative operators in this case.} 

It is not difficult to calculate
\beq
D_0(s) \, =\frac{1}{4\pi}\,  \ln  \left ( {\Lambda^2\over -s}\,+1\,\right )
\,,\qquad s\equiv p^2\,, 
\label{6D0}
\eeq
where $\Lambda^2$ is an ultra-violet cut-off. 
With this expression for $D_0$ the function $G(p^2, 0)$ develops a {\em cut}
on the positive semiaxes of $s$ due to the logarithmic behavior of
$D_0(s)$. This fact has a physical interpretation. Since the
extra dimensions are
noncompact in the model under consideration,
the spectrum of the theory, as seen from the 4D standpoint,
consists of an infinite gapless tower of the KK modes.  
This generates a
cut in the Green's function for $s$ ranging
from zero to $+\infty$.  

In addition, there might exist isolated
singular points  in $G(p^2,0)$. These singularities (for
$s\ll \Lambda^2$) are determined by the equation
\beq
G^{-1}(s,0)\equiv  s - m_c^2 \, \left[
\frac{1}{4\pi}\ln \left(\frac{\Lambda^2}{ -s}\right)\right]^{-1} =\,0 \,,
\label{spoles1}
\eeq
where
$m_c^2$ is defined in Eq.~(\ref{mc}).
Let us introduce the notation
\beq
s\, \equiv \, s_0\,{\rm exp} (i\gamma)\,,
\label{scom}
\eeq
where $s_0$ is a {\em real positive} number. Then,
Eq. (\ref {spoles1}) has two solutions of the form
\beq
s_{0}\approx 4\pi\,   m_c^2 \left[ \ln
\frac{\Lambda^2}{ m_c^2 }
\right]^{-1}\,,
\eeq
and 
\beq
\gamma_1 \, \simeq \,-\,{\pi \over {\rm log} (\Lambda^2/m_c^2) }\,,
~~~\gamma_2 \, \simeq \, 2\pi \,+ \,{\pi \over {\rm log}
(\Lambda^2/m_c^2) }\,.
\label{mpole}
\eeq
We conclude that there are two complex-conjugate poles
on the nearby unphysical Riemann sheets. These poles
cannot be identified with any physical states
of the theory. They are, in fact, manifestations of a 
massive resonance state.  All other complex 
poles appear on subsequent nonphysical Riemann sheets.

\subsection{More than six dimensions}
\label{sevend}

Physics at $D>6$ is similar to that of the six-dimensional world which  
was described in Sect.~\ref{sd}. There are 
minor technical differences between odd- and even-dimensional
spaces, however, as we will discuss momentarily.

In seven dimensions we find
\beq
D_0(s) \,= \,{1\over 2\,\pi^2}\,\left\{ \Lambda \,-
\, \sqrt {-s}\,{\rm arctan}\left( {\Lambda \over \sqrt {-s} } 
\right)  \right\} \,.
\label{7D0}
\eeq
As in the 6D case, there is a branch cut. The cut in this case is due to
the dependence of the Green's function on $\sqrt{s}$.
No other singularities appear on the  physical Riemann sheet.
All  poles are on unphysical Riemann sheets, as previously.

In  the eight-dimensional space the 
expression for $D_0$ reads
\beq
D_0(s) \,= \, {1\over 16\,\pi^2}\,  \left\{       \Lambda^2 \,+\,s\,\left(
{\rm ln} \,  {\Lambda^2 \over -s}\,+ \, 1\,\right) \right\}\,.
\label{8D0}
\eeq
Again, we find a cut due to the logarithm,
 similar to that of the 6D case. All isolated singularities 
appear on unphysical Riemann sheets.

The nine-dimensional formula runs parallel to  that  in 
seven dimensions,
\beq
D_0(s)\,= \,{1\over 12\,\pi^3}\,\left\{ {\Lambda^3 \over 3} \,+\,s
\left( \Lambda \,-
\, \sqrt {-s}\,{\rm arctan} \,  {\Lambda \over \sqrt {-s} } 
 \right) \right\}\,.
\label{9D0}
\eeq
Finally, in ten dimensions 
\beq
D_0(s)\,= \,{1\over 128\,\pi^3}\, \left\{ {\Lambda^4 \over 2}
\, +\, s \left [ \,\Lambda^2 \,+\,s\,\left( 
{\rm ln} \, {\Lambda^2\over -s}\,+ \,1\,\right ) \right ] \right\}\,.
\label{10D0} 
\eeq
The pole structure of $G$ is identical to that of 
the eight-dimensional case.  Since the pattern is now 
well established and clear-cut,
there seems to be no need in  dwelling on higher dimensions. 

Before turning to gravitons we would like to make 
comments concerning the UV cutoff $\Lambda$. 
The crossover distance $r_c\sim m_c^{-1}$ 
depends on this scale:  in 6D the dependence is 
logarithmic, while  in $D>6$ this dependence presents a power law
\cite {DG,Ignatios}. Hence, the crossover scale 
in the $N\ge 2$ DGP models, unlike 
that in the $N=1$ model, is sensitive
to  particular details of the UV completion of the theory.  
Since in the present 
work we adopt an affective  low-energy field-theory strategy,
we are bound to follow the least favorable
scenario in which  the cutoff and the bulk gravity scale 
coincide with each other and both are equal to $\m \sim 10^{-3}$ eV. 
If a particular UV completion were available, it could well
happen that the UV cutoff and bulk gravity scale were different
from the above estimate. In fact, in 
the string-theory-based construction of Ref. ~\cite {Ignatios} the 
UV completion is such
that  the cutoff and bulk gravity scale are in the ballpark of TeV.

In conclusion of this section it is
worth noting that the Green's function $D_0$
in the $N\ge 3$ case contains terms   responsible for 
branch cuts. These terms  are suppressed by  powers of $s/\Lambda $, 
and, naively, could have been neglected.
It is true, though,  that the explicit form of these terms is UV-sensitive
and cannot be established without the knowledge of UV physics.
One  should be aware of these terms since 
they  reflect   underlying physics  -- the presence 
of the infinite tower of the KK states. Fortunately, none of the 
results of the present work depend on these terms.

\subsection{The graviton propagator}
\label{potg}

Now it is time to turn to  gravitons with their specific tensorial
structure. We will consider and analyze the equation of 
motion of the DGP-type model presented in Eq.~(\ref{geq}), which we 
reproduce here again for convenience
\beq
&&\d \, \mpl^2 \, \left ( a\,G^{(4)}_{\mu\nu}+
b\,G^{(D)}_{\mu\nu} \right )
\delta^\mu_A\,\delta^\nu_B
\nonumber\\[3mm]
&&+\, \m^{2+N} \,G^{(D)}_{AB}\,
= \, - T_{\mu\nu}(x)\, \delta^\mu_A\,\delta^\nu_B \, \d \,.
\label{geq1}
\eeq
Here   $G^{(4)}$ and $ G^{(D)}$ 
denote the four-dimensional  and
$D$-dimensional Einstein tensors, respectively,
while $a$ and $b$ are certain constants satisfying the constraint
 $$a+b=1\,.$$
For simplicity we choose    a source  term localized on the brane,
namely,  $T_{\mu\nu}(x)\d $. At the effective-theory level the ratio
$a/b \equiv a/(1-a)$ is a free parameter. The only guidelines
we have for its determination are (i) phenomenological viability; (ii)
intrinsic selfconsistency of the effective theory
which, by assumption, emerges as a low-energy limit of a selfconsistent 
UV-completed underlying ``prototheory." Specifying the
prototheory would allow one to fix the ratio $a/(1-a)$ in terms of fundamental
parameters.

\vspace{1mm}

Our task is to study the gravitational field produced by
the source $T_{\mu\nu}(x)\d $. To this end we 
linearize Eq.~(\ref{geq1}). If $g_{AB}\equiv \eta_{AB}+2h_{AB}$,
in the linearized in $h$ approximation
 we find
\beq
G^{(D)}_{AB}&=&\partial_D^2\,h_{AB}\,-\,\partial_A \,\partial_C\,
h^C_B \, -\,\partial_B \,\partial_C\,
h^C_A \nonumber \\[3mm]
&+& \partial_A\,\partial_B\,h^C_C \, -\, \eta_{AB}\,
\partial_D^2\,h^C_C \,+ \, \eta_{AB}\,\partial_C\,\partial_D\,h^{CD}\,,
\label{GD}
\eeq
where $\partial_D^2 \equiv \partial_D \partial^D$.
On the other hand, the four-dimensional Einstein tensor in the linearized
approximation is
\beq
G^{(4)}_{\mu\nu}&=& \partial_\beta^2\,h_{\mu\nu}\,-\,\partial_\mu \,
\partial_\alpha \,
h^\alpha_\nu \, -\,\partial_\nu  \,\partial_\alpha \,
h^\alpha_\mu \,+\,\partial_\mu\,\partial_\nu\,h^\alpha_\alpha
\,\nonumber \\[3mm]
 &-&  \eta_{\mu\nu}\,
\partial_\beta^2\,h^\alpha_\alpha \,+ \, \eta_{\mu\nu}\,\partial_\alpha
\,\partial_\beta\,h^{\alpha \beta}\,.
\label{G4}
\eeq
In what follows we will work in the harmonic gauge, 
\beq
\partial^A\,h_{AB}\,=\,{1\over 2}\,\partial_B\,h^C_C\,.
\label{har}
\eeq
The advantage of this gauge is that in this gauge the expression 
for $ G^{(D)}_{AB}$ significantly simplifies,
\beq
G^{(D)}_{AB}\,=\,\partial_D^2\,h_{AB}\,-\,{1\over 2}\,
\eta_{AB}\,\partial_D^2\,h^C_C \,.
\label{GDs}
\eeq
Additional conditions which are invoked to solve the
$\{ab\}$ and $\{a\mu\}$ components of the equations of motion are
\beq
h_{a\mu}=0,~~~~~h_{ab}\,=\,{1\over 2}\,\eta_{ab}\,h^C_C\,.
\label{abp}
\eeq
Using the last equation it is not difficult to obtain the relation
\beq
N\,h^\mu_\mu\,=\,(2-N)\,h^a_a \,.
\label{mua}
\eeq
This relation obviously suggests that we should consider separately 
two cases: 

(i)   $N=2$;

 (ii)   $N>2$.\\
 We will see, however, that
the results in the  $N=2$ and $N>2$ cases are somewhat similar.

\subsection{Brane-induced gravity in six dimensions ({$N=2)$}}
\label{igisd}

In two extra dimensions  Eq. (\ref {mua}) implies
\beq
h^\mu_\mu\,=\,0\,.
\label{mumu0}
\eeq
Therefore, the trace of the $D$-dimensional graviton
coincides with the trace of the extra-dimensional part,
\beq
h^A_A\,=\,h^a_a\,.
\label{aaaa}
\eeq
As a result, the four-dimensional components
of the harmonic gauge condition (\ref {har})
reduce to
\beq
\partial^\mu \,h_{\mu\nu}\,=\,{1\over 2}\,\partial_\nu\,h^a_a\,.
\label{4dgauge}
\eeq
Let us now have a closer
look at the $\{\mu\nu \}$ part of Eq. (\ref {geq}).
Taking the trace of  this equation and using Eqs.~(\ref {GDs}), 
(\ref {G4}),
(\ref {mumu0}) and (\ref {4dgauge})
we arrive at\footnote{As before, we put the 
transverse derivatives to be zero in the $\Delta\to 0$ limit.}
\beq
(3b\,-\,1)\,\d \, \mpl^2 \, \partial^2_\mu \,h^a_a\,+\,
2\,\m^{2+N} \,\partial^2_A \, h^a_a\, = \, T^\mu_\mu\, \d \,.
\label{6Dtrace}
\eeq
The  obtained  equation is very similar to the scalar-field 
equation (\ref {seq}). Therefore, we will follow the same route 
as in  the scalar-field case,
until we come to a subtle point, a would-be obstacle,
which was nonexistent  in the scalar-field case. 

Let us Fourier-transform  Eq. (\ref {6Dtrace}),
\beq
&&(3b\,-\,1) \,\d \, \mpl^2 (-p^2)\, {\tilde h}^a_a(p,y) 
\nonumber\\[3mm]
&& + 2
\m^{2+N} \,(-p^2 -  \Delta_y )\, {\tilde h}^a_a(p,y) =
{\tilde T}(p)\, \d \,.
\label{6Dtracemom}
\eeq
The general solution of the above equation is
\beq
&& {\tilde h}^a_a(p,y)=   {{\tilde T}(p)\over \mpl^2}\,\g (p,y) \,,
\label{defg}\\[4mm]
&& \g = { D(p,y) \over 2m_c^2 \,-\,(3b\,-\,1)  p^2\, D_0(p)\,}\,
+\,c\, \g_{\rm hom}\, ,
\label{g6D}
\eeq
where  the solution of the homogeneous equation takes the form
\beq
\g_{\rm hom}\,=\, D(p,y)\,\delta \left (2m_c^2 \,-\, (3b\,-\,1) p^2D_0(p) 
\right )\,.
\label{hom6D}
\eeq
To begin with, let us consider  the case  $3b>1$. Then
the first term on the right-hand side  of Eq.~(\ref {g6D}) has  poles
for complex values of  $p^2$. For instance, in the 6D case
this pole is determined by the equation
\beq
s= {2m_c^2\over (3 b - 1) D_0(s)}= {4\pi \,{2m_c^2}\over (3b-1)}\, 
\left[\ln \frac{\Lambda^2}{-s} \right]^{-1}\,.
\label{6Dnewpoles}
\eeq
This equation has at least two solutions of the form
\beq
s_{*}\approx {4\pi\,   2m_c^2 \over 3b-1}
\left[ \ln \frac{\Lambda^2}{ m_c^2 }
\right]^{-1}\,,
\label{som}
\eeq
and 
\beq
\gamma_1 \, \simeq \,-\,{\pi \over {\rm log} (\Lambda^2/m_c^2) }\,,
~~~\gamma_2 \, \simeq \, 2\pi \,+ \,{\pi \over {\rm log}
(\Lambda^2/m_c^2) }\,.
\label{*pole}
\eeq
The quantity ${\tilde h}^a_a(p,y) $ is not a gauge invariant variable. 
Therefore, the presence  of certain poles in the expression for
${\tilde h}^a_a(p,y)$ depends on a gauge. However, explicit 
calculations (see below) show that the poles found above 
also enter the gauge invariant physical amplitude. 
Therefore, we need to take these poles seriously and analyze their
physical consequences.

\subsection{$b>1/3$}
\label{2bba}

If $b>1/3$  there are no poles on the physical Riemann sheet.
Instead,   poles appear on the nearest unphysical 
Riemann sheets. These poles cannot be identified with any physical states
of the theory. They represent a signature of  
massive resonance states.  All other complex  poles appear on 
subsequent nonphysical Riemann sheets.

Using a contour integral one can easily write down the spectral representation 
for the Green's function ${\mathcal G}$
\beq
{\g}(p,y=0) \,= \, {1\over \pi}\,\int_0^{\infty}\,
{ \rho (t) \,dt \over \,t\,- \,p^2\,-\,i\,{\bar \e}}\,,
\label{disp4D}
\eeq 
where the spectral function is defined as 
\beq
\rho (t)\,=\,{2\,m_c^2\, {\rm Im}\, D_0(t) 
\over\left[ (3b-1) t\,{\rm Re}\, D_0\,-\,2m_c^2\right]^2\,+\left[(3b-1)t\,
{\rm Im}\, D_0)\right]^2}\,,
\label{rho}
\eeq 
and  
\beq
{\rm Im}D_0\,=\,\pi\,\int \,{d^N q\over (2\pi)^N}\,\delta(t-q^2)\,
=\,{\pi^{N+2\over 2} \over (2\pi)^N \Gamma(N/2)}\,
t^{N-2\over 2}\,.
\label{ImD0}
\eeq 
We see that $\rho(t)$ satisfies the positivity requirement.
Equation (\ref {disp4D}) guarantees that the Green's 
function $ \g $ is causal.

The next  step is applying the expression 
for $\g$ to   calculate ${\tilde h}_{\mu\nu}$. In fact, it is more convenient 
to calculate the tree-level amplitude
\beq
A(p,y)\,\equiv \,{\tilde h}_{\mu\nu}(p,y)\,{\tilde T}^{\prime\,\mu\nu}
(p)\,,
\label{amp}
\eeq
where ${\tilde T}^{\prime\,\mu\nu}(p)$ is a
conserved energy momentum tensor,
 $$
 p_\mu\,{\tilde T}^{\prime\,\mu\nu}\,
=\,p_\nu\,{\tilde T}^{\prime\,\mu\nu}\,=\,0\,.
$$
Using Eqs. (\ref{geq1}),  (\ref {defg}) and  (\ref{6Dc}) 
we obtain a lengthy expression for the amplitude $A(p,y)$,
\beq
A(p,y)&=&{1\over \mpl^2}\,{D(p,y)\over p^2\,D_0(p)\,-\,m_c^2}\,
\left \{{\tilde T}_{\mu\nu}  {\tilde T}^{\prime\,\mu\nu}
\right.\nonumber\\[3mm]
&-&\left.
{{\tilde T}\, {\tilde T}^{\prime}   \over 2}\,
\left[ {(2b-1)p^2D_0 -m_c^2   \over (3b-1)p^2D_0 - 2m_c^2 }  
  \right] \right \} .
\label{a}
\eeq

\vspace{2mm}

\noindent
Let us study  the above expression in some detail. The first
question to ask is about poles.
It is quite clear that the $p^2$-poles of
 $A$ are of two types; their position is determined by:
$$
p^2\,D_0(p)\,=\,m_c^2
$$ 
or
$$
(3b-1) p^2\,D_0(p)\,=\,2m_c^2\,.
$$
 As was explained previously, 
all these poles appear on the second  Riemann sheet,
 with the additional  images on other unphysical sheets.
None of these poles can be identified with asymptotic physical 
states. As was elucidated above, the occurrence
of the poles on the second and subsequent Riemann sheets corresponds
to the massive-resonance nature of the effective 4D graviton.
Our previous analysis can be repeated practically {\em verbatim},
 with minor modifications, 
proving analyticity and causality of the amplitude $A$.

Next, we observe that at  large momenta, i.e., 
when $p^2\,D_0(p)\gg m_c^2$, the scalar part of 
the propagator has 4D behavior;  the tensorial structure 
is not four-dimensional, however. 
The   terms in the   braces  in Eq.~(\ref {a}), namely, 
\beq
\left \{{\tilde T}_{\mu\nu}  {\tilde T}^{\prime\,\mu\nu}\,-\,
{2b-1\over 2(3b-1)}\, {\tilde T}\, {\tilde T}^{\prime}\,\right \}\,,
\label{2polar}
\eeq
correspond to the exchange of massive gravitons and
scalar degrees of freedom. This would give rise to additional 
contributions in the light bending, and is excluded phenomenologically, 
unless the contribution  due to  extra polarizations is canceled 
by some other interactions (such as, e.g.,  an additional  
repulsive vector exchange). Note also that when $b\gg a$, i.e., 
$b\to 1$, one obtains the 
tensorial structure of  6D gravity, as expected from (\ref {11}).

On the other hand, at large distances,
i.e., at $p^2\,D_0(p)\ll m_c^2$, we get 
the following tensorial structure of the amplitude (\ref {a}): 
\beq
\left \{{\tilde T}_{\mu\nu}  {\tilde T}^{\prime\,\mu\nu}\,-\,
{1\over 4}\, {\tilde T}\, {\tilde T}^{\prime}\,\right \}\,.
\label{6Dpolar}
\eeq
This exactly  corresponds to the exchange of a six-dimensional 
graviton, as was expected.

\subsection{$b<1/3$}
\label{2bma}

This case is conceptually different from that of Sect.~\ref{2bba}.
As we will see momentarily, if $b<1/3$ there are no
problems in (i) maintaining 4D unitarity; and (ii)
getting the appropriate 4D  tensorial structure of gravity
at subhorizon distances. This is achieved at a price of
abandoning 4D analyticity, in its standard form,
which could presumably lead to the loss of causality
at distances of the order of $m_c^{-1}\sim 10^{28}$ cm. 
The absence of  causality at distances   $ \gsim 10^{28}$ cm,
was argued recently \cite{ADDG} to be an essential 
ingredient for solving the cosmological constant problem.

Although all derivations and conclusions are quite similar for
any ratio $a/b$ as long as $2b<a$, we will stick to the technically
simplest example  $b=0$, $a=1$.
In the situation at hand,  the homogeneous 
part (\ref {hom6D}) need not
be trivial, i.e $c$ need not vanish.
The value of the constant $c$ is  determined once 
the rules for the pole at $ p^2D_0(p) + 2m_c^2 =0 $ are specified.
Putting $c=0$  leads
to  {\em nonunitary} Green's function.
Therefore, we abandon the condition  $c=0$ in an attempt to make
a more consistent choice that would guarantee 4D unitarity.
We stress that we are after unitarity here, not unitarity plus causality.

To begin with we pass to the Euclidean space in $p^2$ (i.e. $p^2\to p_E^2$)
and introduce the following notation:
\beq
P^{(E)}(p_E^2)\,\equiv\, {1\over 2\,m^2_c\,-\,p_E^2\,D_0(p_E)-\,i\e}\,.
\label{PE}
\eeq
The function $P^{(E)}$    
is  a Euclidean-space solution of Eq. (\ref {6Dtracemom}), 
with the particular choice $c=i\pi$. (The choice 
$c=-i\pi$ would lead to Eq.~(\ref {PE}) with the 
replacement  $\e \to -\e$). 

As the next  step we will analyze  the complex plane of $p_E^2$.
Since the function $D_0(p_E)$ is real, the function
$P^{(E)}(p_E^2)$ must (and does) have an isolated singularity in the $p_E^2$ plane
 which is similar to a conventional massive pole, except that it lies
 in the Euclidean domain. 
This singularity occurs at the point $p_E^2=p_*^2$  
is defined by the condition
\beq
p_*^2D_0(p_*)\,=\, 2\,m_c^2\,,~~~~~p_*^2\,\,\,\mbox{real and positive}. 
\label{pstar}
\eeq
This is the only isolated singularity in Eq.~(\ref {PE});
it is located in the complex  $p_E^2$ plane 
on the real positive semiaxis. In addition to this pole
singularity, the 
function (\ref {PE}) has a branch cut stretching from zero to
$-\infty$ due to the imaginary part of $D_0(p_E^2)$ 
appearing at negative values of $p_E^2$.
As before, this branch cut is the reflection
of an infinite gapless tower of the KK states.  
As a result, the following spectral representation obviously emerges
 for $P^{(E)}(p_E^2)$:
\beq
P^{(E)}(p_E^2+i\e)\,=\, 
{1\over \pi}\,\int_0^{-\infty}\,
{{\rm Im}P^{(E)}(u)\,du \over u\,-\,p_E^2\,}\,
+\, {R\over p_*^2 \,-\,p_E^2 \,-\,i\e}\,,
\label{PEdisp}
\eeq
with the Euclidean pole term being ``unconventional."
The residue of the pole $R$ is given (for any $N$) by
\beq
R^{-1}\,=\,\int\, {d^Nq\over (2\pi)^N}\,{q^2\over (q^2\,+\, p_*^2)^2}\,.
\label{R}
\eeq

Note that in the first term on the right-hand side 
of Eq.~(\ref {PEdisp}) the integration  runs from zero to minus infinity; thus,
the integrand never hits the would-be pole at $u=p_E^2>0$. 
Therefore, the $i\e$ prescription is in fact used only to specify the 
isolated pole at $p_E^2=p_*^2$.

We proceed further and define a {\it symmetric} function
\beq
\Pi^{(E)}(p_E^2)\, \equiv \, { 1 \over 2} \left\{
P^{(E)}(p_E^2-i\e)\,+\,P^{(E)}(p_E^2+i\e) \right\}\,.
\label{PiE}
\eeq
It is just this symmetric function on which we will focus
in the remainder of the section.
Let us return to the Minkowski space. This is done by substituting
$$
 p_E^2 \to {\rm exp}(-i\pi)p^2\,,\qquad  u\to {\rm exp}(-i\pi) t
 $$
in Eq.~(\ref {PEdisp}). Furthermore, observing that
${\rm Im}\, P ={\rm Im}\, \Pi $, we obtain the following
representation for the Minkowskian $\Pi$:
\beq
\Pi (p) \,=\,{1\over \pi}\,\int_0^{\infty}\,
{{\rm Im}\Pi (t)\,dt \over t\,-\,p^2\,-\,i\,{\bar \e}}\, +\,\Pi_0(p)\,,
\label{Pi}
\eeq
where 
\beq
\Pi_0(p)\,\equiv \,{1\over 2}   \left ( 
{R\over p_*^2 \,+p^2 \,-\,i\e}\,+\,
{R\over p_*^2 \,+p^2 \,+\,i\e} \right )\,.
\label{Pi0}
\eeq
It is necessary to emphasize
 that $\epsilon $ and $\bar\epsilon $ are two distinct regularizing
parameters, $\epsilon\neq\bar\epsilon$. The parameter $\epsilon$ is used 
to regularize the pole at $p^2=-p_*^2$, while $\bar\epsilon$
sets the rules for the branch cut.
The most important property of $\Pi$    is 
that the pole at $p^2=-p_*^2$ has no imaginary  part, by construction. Hence, 
there is no physical particle that corresponds to this pole.  
In the conventional local field theory the only 
possible additions with no imaginary part are polynomials.
Here we encounter a new structure which will be discussed in more detail
at the end of this section. 

Our goal   is to show that a 4D-unitarity-compliant  spectral 
representation holds
for the Green's function on the brane, at least in the domain where 
the laws of 4D physics are applicable. To this end we turn to the 
function $\g(p,y)$,   defined as  
\beq
\g \,=\, {D(p,y)}\,\Pi(p^2)\,.
\label{6Dc}
\eeq
with the purpose of studying its properties. 
It is convenient to pass to the momentum space with respect to 
extra coordinates too. Then, the propagator (\ref {6Dc}) 
takes the form
\beq
{\tilde \g}(p,q)\,=\,{\Pi(p^2)\over q^2 \,-\, p^2\, - i\,{\bar \e}}\,.
\label{pq}
\eeq
With these definitions in hand, we can write down 
the 4D  dispersion relation. We start from  the 
K\"allen--Lehman representation for the propagator (\ref {pq}).
As we will check below, this representation takes the form 
\beq
{\tilde \g}(p,q)\,= \, {1\over \pi}\,\int_0^{\infty}\,
{{\rm Im}{\tilde \g}(t,q) \,dt \over t\,-\,p^2\,-\,i\,{\bar \e}}\,
+\,{\Pi_0(p^2) \,-\,\Pi_0(q^2) \over q^2\, -\,p^2\,- 
\,i\,{\bar \e}}\,.
\label{disp}
\eeq
The imaginary part of ${\tilde \g}$ is defined as follows
\beq
{\rm Im}{\tilde \g}(t, q)\,=\,\pi \,\delta (q^2\,-\,t)\, 
{\rm Re}\Pi(t)\,+\,
{\rm Im}\Pi(t)\,{\mathcal P}{1\over q^2\,-\,t}\,,
\label{Im}
\eeq
where ${\mathcal P}$ stands for the {\it principle value} of a singular function,
\beq
{\mathcal P} \, {1\over q^2\,-\,t}\, = \,{1\over 2}
\left ( {1\over q^2\,-\,t\,+\,i\delta} \, + \,
 {1\over q^2\,-\,t\,-\,i\delta}\right ). 
\label{P}
\eeq
The fact that Eq. (\ref {disp}) holds can be checked by substituting 
(\ref {P}) and (\ref {Im}) into (\ref {disp}) 
and exploiting the relation
\beq
&&{1\over \pi}\,\int_0^{\infty}\,
{{\rm Im}{\Pi}(t) \over t\,-\,p^2 \,-\,i\,{\bar \e}}\,
{\mathcal P}{1\over q^2\,-\,t }\,dt 
\nonumber\\[3mm]
&&=\,- 
{{\rm Re}\Pi(q^2) \,-\,\Pi_0(q^2)\,+\,\Pi_0(p^2)\,-\,\Pi(p^2)
\over  q^2-p^2 - i\,{\bar \e}}\,.
\label{check1}
\eeq
This turns Eq. (\ref {disp}) into  identity.

Finally we approach  the main point  of this  section -- 
the dispersion relation for $\g (p,y=0)$,
 the Green's function on the brane.
As such, it must have a spectral representation with the positive 
spectral density, as we have already seen from the KK-based analysis. 
The positivity is in one-to-one correspondence with the 4D unitarity.

The dispersion relation can be obtained by integrating
(\ref {disp}) with respect to $q$,
\beq
{\g}(p,y=0)\,= \, {1 \over \pi}\,\int_0^{\infty}\,
{ \rho (t) \,dt \over \,t\,- \,p^2\,-\,i\,{\bar \e}}\,
+\,\Pi_0(p^2)\,{\rm Re}D_0(-p_*^2)\,.
\label{disp4Dp}
\eeq 
According to Eq. (\ref {disp}), the spectral density 
$\rho$ is defined as
\beq
\rho (t)\,=\,\int \,{d^N q\over (2\pi)^N}\,
{\rm Im}\, {\tilde \g}(t, q)\,.
\label{rhoDEF}
\eeq 
The first term on the right-hand side
in Eq.~(\ref{disp4Dp}) is conventional while the second is not,
and we hasten to discuss it. This term has no imaginary part, by construction.
Hence, it does not contribute to the unitarity cuts
in diagrams. Therefore, this term does not affect the 
spectral properties. 

As was mentioned, in conventional 4D theories 
only a finite-order polynomial in 
$p^2$ that has no imaginary part can be added to or subtracted from 
the dispersion relation. This is because, normally one  deals
 with Lagrangians which contain only a finite 
number of derivatives, i.e., a finite number of terms  
with positive powers of  $p^2$ in the momentum space.
In the problem under consideration  this is not the case, however.
In fact, no  local 4D Lagrangian exists in our model 
 at all,  and yet we are studying the spectral 
properties in terms of the intrinsically 4D variable,  $p^2$. 
The theory (\ref {1}) is inherently higher-dimensional because
of the infinite volume of  the extra space. One can try to
``squeeze" it in four dimensions at a price of having 
 an  {\it infinite} number of 4D fields.  
For such a theory  there is no  guarantee that  
{\it analyticity} of the Green's functions 
in terms of the 4D variable $p^2$ will hold because 
 the  effective  4D Lagrangian
obtained by ``integrating out'' the infinite gapless 
KK tower  will necessarily contain \cite{DGS} nonlocal terms of the
type  $\partial^{-2}$. (Note that a similar prescription  for the poles 
in a pure 4D local theory \cite {Lee} is hard to reconcile with 
the path integral formulation \cite {Gross}. In our  case this is not 
a  concern  since the theory is not local in four-dimensions 
in the first place.)

Therefore, it is only natural that 4D unitarity can be maintained but 4D 
analyticity cannot. Nonanalyticity 
leads to violation of causality, generally speaking. That is to say, the 
Green's function (\ref {disp4Dp}) is acausal. 
Therefore, we have an apparent violation of causality 
in the 4D slice of the entire $(4+N)$ dimensional theory
which, by itself, {\em is causal}. The apparent acausal effects can 
  manifest  themselves only at the  scale  of the order of
$m^{-1}_c\sim 10^{28}$ cm. In fact, as was noted in \cite{ADDG},
this is a welcome feature  for a possible solution of the cosmological 
constant problem.

Let us now return to the first term on the right-hand side
of Eqs. (\ref {disp4Dp}). Using Eqs.~(\ref {Pi}) 
and (\ref {Im}) we can calculate
the spectral function which comes out as follows:
\beq
\rho (t)\,=\,{2\,m_c^2\, {\rm Im}D_0(t)\over (t\,{\rm Re}D_0\,+
\,2m_c^2)^2\,+\,(t\,{\rm Im}D_0)^2}\,,
\label{rhop}
\eeq 
where 
\beq
{\rm Im}D_0\,=\,\pi\,\int \,{d^N q\over (2\pi)^N}\,\delta(t-q^2)\,
=\,{\pi^{N+2\over 2} \over (2\pi)^N \Gamma(N/2)}\,
t^{N-2\over 2}\,.
\label{ImD0p}
\eeq 
We see that $\rho(t)$ satisfies the positivity 
requirement.\,\footnote{For $N\ge 5$ 
the integral in Eq.~(\ref {disp4Dp}) diverges. However, since 
our model has a manifest UV cutoff $\Lambda$, the above integral 
must be cut off at $\Lambda$. Alternatively, one could use
a dispersion relation with subtractions.}
 
\vspace{1mm}

Next we observe that at  large momenta, i.e., 
at $p^2\,D_0(p)\gg m_c^2$, the propagator we got
has the desired 4D behavior. For the scalar part of the propagator this 
is expected.
However, with regards to the tensorial structure 
this circumstance is less trivial.
If $p^2\,D_0(p)\gg m_c^2$ the   terms in the  braces in Eq.~(\ref {a}), 
\beq
\left \{{\tilde T}_{\mu\nu}  {\tilde T}^{\prime\,\mu\nu}\,-\,
{1\over 2}\, {\tilde T}\, {\tilde T}^{\prime}\,\right \}\,,
\label{2polarp}
\eeq
correspond to the exchange of two physical graviton polarizations.
 Therefore, for the observable 
distances  the tensorial structure of the massless
4D graviton (\ref {2polarp}) is recovered. 

On the other hand, for large  (super-horizon) distances,
 $p^2\,D_0(p)\ll m_c^2$, we get 
a different tensorial structure of the same  amplitude,
\beq
\left \{{\tilde T}_{\mu\nu}  {\tilde T}^{\prime\,\mu\nu}\,-\,
{1\over 4}\, {\tilde T}\, {\tilde T}^{\prime}\,\right \}\,.
\label{6Dpolarp}
\eeq
This exactly corresponds to the exchange of the six-dimensional 
graviton.

\subsection{{$D>6$}}
\label{dls}

Corresponding calculations and results are quite similar to the
$D=6$ case, with minor technical distinctions which we summarize below.
For $N\neq 2$
\beq
h_{ab}\,=\,{1\over 2\,-\,N}\,\eta_{ab}\,h^\mu_\mu\,.
\label{hab}
\eeq
Therefore,  we get
\beq
\partial^\mu \,h_{\mu\nu}\,=\,{1\over 2\,-\,N}\,\partial_\nu\,
h^\alpha_\alpha\,.
\label{hmn}
\eeq
Then, the trace of the $\{\mu\nu\}$ equation takes the form
\beq
&& k_N\, \d \, \mpl^2 \,
  \partial^2_\mu \,h^a_a\,+\,
\m^{2+N} \,(N+2)\, \partial^2_A \, h^a_a
\nonumber\\[3mm]
 =&& {N} \, T^\nu_\nu\, \d \,.
\label{traceeqD}
\eeq
where $k_N\equiv 2-N(2-3b)$. The above equation 
which  can be used to find the solution we are after. 
We proceed parallel to  the 
six-dimensional case.  Let us introduce the notation
\beq
{\tilde h}^a_a(p,y)\, = \,{N}\, 
{{\tilde T}(p)\over \mpl^2}\,\g_N (p,y) \,,
\label{defgN}
\eeq
where 
\beq
\g_N = { D(p,y) \over - k_N\,p^2\,D_0(p) +  u^N\,(N+2)}
+\,c\, \g_{\rm N\,hom}\, .
\label{gD}
\eeq
The solution of the homogeneous equation takes the form
\beq
\g_{\rm N\, hom}\,=\, D(p,y)\,\delta \left (-k_N\,p^2\,D_0(p) \,+\, u^N
\,(N+2) \right )\,.
\label{homD}
\eeq
Here  $$u^N\equiv \m^{2+N}/\mpl^2\,.$$
As in the 6D case, we conclude that that there exists a 
solution to the equation
$$
-k_N p^2\,D_0(p) \,+\, u^N(N+2)  =0
$$  
with a 
complex value of $p^2$. These poles occurs on the nonphysical sheets
as long as $k_N>0$, so the Green's function admits the spectral representation.

Using the expressions above one readily calculates  
the tree-level amplitude $A$,
\beq
A(p,y)&=& {1\over \mpl^2}\,{D(p,y)\over p^2\,D_0(p)\,-\,
u^N }\nonumber \\[3mm]
&\times & 
\left \{{\tilde T}_{\mu\nu}  {\tilde T}^{\prime\,\mu\nu}\,-\,
{ {\tilde T}\, {\tilde T}^{\prime}  \over 2}\,
\left ( {(k_N-bN)p^2 D_0 - 2u^N  \over k_N p^2 D_0 - (2+N)u^N}    \right )
\right \}\,.
\label{aD}
\eeq

\subsection{$b> (2N-2)/3N$}
\label{2bbap}

In this case there are no poles on the physical Riemann sheet.
Hence, all the poles are of the resonance type. The tensorial structure 
at large distances is that  of the D-dimensional theory
\beq
\left \{{\tilde T}_{\mu\nu}  {\tilde T}^{\prime\,\mu\nu}\,-\,
{1\over 2+N}\, {\tilde T}\, {\tilde T}^{\prime}\,\right \}\,.
\label{DpolarD}
\eeq 
However, the tensorial structure at short distances $\lsim m_c^{-1}$
differs from that of 4D massless gravity. Hence, some additional 
interactions, e.g., repulsion due to a vector field, 
is needed to make this theory consistent with   data.

\subsection{$b< (2N-2)/3N$}
\label{2bmap}

The consideration below is very similar to the 6D case.
In perfect  parallel with the 6D case we consider for simplicity 
only the $b=0$ case and define the function
\beq
P_N^{(E)}(p_E^2)\,\equiv\, {1\over u^N(N+2)/(2N-2) \,-\,
 p_E^2\,D_0(p_E)-\,i\e}\,,
\label{PEN}
\eeq
which  has a spectral representation:
\beq
P_N^{(E)}(p_E^2+i\e)\,=\, 
{1 \over \pi}\,\int_0^{-\infty}\,
{{\rm Im}P_N^{(E)}(u)\,du \over u\,-\,p_E^2\,}\,
+\, {R\over p_*^2 \,-\,p_E^2 \,-\,i\e}\,.
\label{PENdisp}
\eeq
The residue $R$ is determined by Eq. (\ref {R})
while  $p_*^2$   is now  a solution to the equation
\beq
p_*^2D_0(p_*)\,=\, {u^N (N+2)\over 2(N-1)}\,,~~~~~p_*^2>0\,. 
\label{pstarN}
\eeq
As in the 6D situation, we use the expression (\ref {PENdisp}) to define 
a {\it symmetric} function 
\beq
\Pi^{(E)}_N(p_E^2)\, \equiv \, { 1 \over 2} \left\{
P^{(E)}_N(p_E^2-i\e)\,+\,P^{(E)}_N(p_E^2+i\e) \right\}\,.
\label{PiNE}
\eeq
The latter, being continued to the Minkowski space, admits the following 
spectral representation:
\beq
\Pi_N (p) \,=\,{1\over \pi}\,\int_0^{\infty}\,
{{\rm Im}\Pi (t)\,dt \over t\,-\,p^2\,-\,i\,{\bar \e}}\, +\,\Pi^{(N)}_0(p)\,,
\label{PiN}
\eeq
where 
\beq
\Pi^{(N)}_0(p)\,\equiv \,{1\over 2}   \left ( 
{R\over p_*^2 \,+p^2 \,-\,i\e}\,+\,
{R\over p_*^2 \,+p^2 \,+\,i\e} \right )\,.
\label{Pi0N}
\eeq
As previously,  $\epsilon $ and $\bar\epsilon $ are 
two {\em distinct} regularizing parameters, $\epsilon\neq\bar\epsilon$.

For the Green's function of interest  
\beq
\g_N \,=\, {{D(p,y)}\,\Pi_N(p^2)\over 2N-2} 
\label{Dc}
\eeq
we repeat  the analysis of Sects.~\ref{igisd},  
\ref{2bba} and \ref{2bma}  to confirm
with certainty that the function $\g_N(p,y=0)$ 
does admit the spectral representation
(\ref {disp4Dp}), with a positive spectral function, 
similar to the 6D case, see Eq.~ (\ref {rhop}). 

The  expression in Eq.~(\ref {aD}) interpolates between
the four-dimensional and D-dimensional patterns.
This was already established for the scalar 
part of the amplitude. Let  us examine  the 
tensorial part. For $p^2\,D_0(p)\gg u^N$ we get 
\beq
\left \{{\tilde T}_{\mu\nu}  {\tilde T}^{\prime\,\mu\nu}\,-\,
{1\over 2}\, {\tilde T}\, {\tilde T}^{\prime}\,\right \}\,.
\label{4DpolarD}
\eeq
This corresponds to two helicities of the 4D massless graviton.
In the opposite limit, $p^2\,D_0(p)\ll u^N$, we recover 
the tensorial structure corresponding to the $(4+N)$-dimensional 
massless graviton.

\subsection{No strong coupling problem}

We start from a brief review of the well-known
phenomenon -- the breakdown of perturbation theory for
the graviton with  the hard mass \cite {PF}, occurring at the
 scale   lower that the UV cutoff of the theory
\cite{Arkady,DDGV,AGS}. We then elucidate as to how
this problem is avoided in the models (\ref{1}),  (\ref{11}).

The 4D action of a massive graviton is
\beq
S_m\,=\,{\mpl^2\over 2} \,\int\,d^4x\,\sqrt{g}\,R(g)\,+ 
\,{\mpl^2\,m_g^2 \over 2}\,
\int\,d^4x\,\left (h_{\mu\nu}^2\,-\,(h^\mu_\mu)^2\right )\,,
\label{PF}
\eeq
where $m_g$ stands for the graviton mass and $h_{\mu\nu}\equiv
(g_{\mu\nu} -\eta_{\mu\nu})/2$. The mass term has the Pauli--Fierz form
  \cite {PF}. This is the only Lorentz invariant form of the
mass term which in the quadratic order in $h_{\mu\nu}$
does not  give rise to ghosts \cite {Neu}.
Higher powers in  $h$  could be  arbitrarily
added to the mass term  since there is
no principle, such as reparametrization
invariance, which could fix these terms.
Hence, for definiteness, we assume that
the  indices in the mass term are raised and lowered by
$\eta_{\mu\nu}$. Had we used  $g_{\mu\nu}$ instead,
the difference would appear only in   cubic and higher
orders in $h$,  which are not fixed anyway.

In order to reveal  the origin of the problem
let us have a closer look at the  free graviton
propagators in the massless and massive theory.
For the massless graviton we find
\beq
D^0_{\mu\nu ;\alpha\beta}(p)\,=\, \left(
{1\over 2} \,{\bar \eta}_{\mu\alpha}  {\bar \eta}_{\nu\beta}+
{1\over 2} \, {\bar \eta}_{\mu\beta}  {\bar \eta}_{\nu\alpha}-
{1\over 2} \, {\bar \eta}_{\mu\nu}  {\bar \eta}_
{\alpha\beta}\right)\frac{1}{
-\,p^2\,-\,i\epsilon}\,,
\label{4D}
\eeq
where
\begin{equation}
\bar\eta_{\mu\nu}\,\equiv \,\eta_{\mu\nu}\,-\,
\frac{p_\mu p_\nu}{p^2}\,.
\label{singp}
\end{equation}
The momentum dependent parts of the tensor structure were
chosen in a particular gauge   convenient for our discussion.
On the other hand, there is no gauge freedom for the
massive gravity presented  by the action (\ref{PF}); hence, the
corresponding propagator is unambiguously determined,
\beq
D^m_{\mu\nu ;\alpha\beta}(p)&=&
\left(
{1\over 2} \,\tilde\eta_{\mu\alpha} \tilde \eta_{\nu\beta}+
{1\over 2} \, \tilde\eta_{\mu\beta}  \tilde\eta_{\nu\alpha}-
{1\over 3} \, \tilde\eta_{\mu\nu} \tilde 
\eta_{\alpha\beta}\right)
\nonumber\\[3mm]
&\times&
\frac{1}{
\,m_g^2\,-\,p^2\,-
i\epsilon}\,,
\label{5D}
\eeq
where
\begin{equation}
\tilde\eta_{\mu\nu}\,\equiv \,\eta_{\mu\nu}\,-\,
\frac{p_\mu p_\nu}{m_g^2}\,.
\label{singm}
\end{equation}
We draw the reader's attention to
 the $1/m_g^4$, $1/m_g^2$ singularities of the above propagator.
The fact of their occurrence  will be important in what follows.

It is the difference in the  numerical coefficients in front of the 
$\eta_{\mu\nu}
\eta_{\alpha\beta}$ structure in the massless vs. massive propagators 
(1/2 versus 1/3) that  leads to the famous perturbative
discontinuity \cite{Iwa,Veltman,Zakharov}.
No matter how small the graviton mass
is, the predictions are substantially  different
in the two cases. The structure (\ref {5D}) gives
rise to contradictions with observations.

However, as was first pointed out in Ref.~\cite {Arkady},
this discontinuity could be an artifact of relying on the
tree-level perturbation theory which, in fact,
badly breaks down at a  higher nonlinear level  \cite {Arkady,DDGV}.
One should note that the discontinuity does no appear on curved backgrounds
\cite{KoganAdS,PorratiAdS} 
-- another indication of the spurious nature
of the ``mass discontinuity phenomenon." 

To see the failure of the perturbative expansion
in the Newton constant $G_N$ one could examine the Schwarzschild
solution of the model (\ref {PF}), as  was done in Ref.~\cite{Arkady}. 
However, probably the easiest way  to 
understand the 
perturbation theory breakdown  is through examination of 
 the tree-level trilinear graviton vertex diagram. 
At the nonlinear level we have two
extra propagators which could provide a singularity in $m_g$ up to 
$1/m_g^8$.

Two leading terms, $1/m_g^8$ and $1/m_g^6$, do not contribute  \cite{DDGV},
so  that the worst  singularity is $1/m_g^4$.
This is enough to  lead to the perturbation theory breakdown. For a
Schwarzschild source of mass $M$ the breakdown  happens \cite{Arkady,DDGV} at
the scale 
$$
\Lambda_m\sim m_g (Mm_g/\mpl^2)^{-1/5}\,.
$$  
The result can also be understood in terms of
interactions of longitudinal polarizations of the massive graviton which 
become strong \cite{AGS}. For the gravitational sector 
{\em per se}, the corresponding scale $\Lambda_m$  reduces to \cite{AGS} 
$$
m_g  (m_g/\mpl)^{-1/5}\,.
$$
If one uses the freedom associated with possible addition of  higher nonlinear
terms,  one can make \cite {AGS} the breaking scale  as large as 
$$
m_g/ (m_g/\mpl)^{1/3}\,.
$$

Summarizing, in the diagrammatic language  the reason
for the precocious breakdown of perturbation theory
can be traced back to the infrared terms in the propagator (\ref {5D}) 
which  scale as
\beq
{p_\mu \,p_\nu\over m_g^2} \,.
\label{singPF}
\eeq
These terms do not manifest themselves at the linear level;
however,  they do contribute to nonlinear vertices creating problems
in the perturbative treatment of massive gravity already in a classical theory.

We will see momentarily  that   similar problems are totally absent
in the propagator of the model (\ref {1}). For illustrational purposes
it is sufficient to treat the $N=2$ case.  All necessary
calculations were carried out  in  
Sect.~\ref{potg}. Therefore,  here we just assemble  relevant answers. 

For $N=2$  and $b>1/3$ we find
\beq
{p_\mu \,p_\nu\,D(p,y) \over 2m_c^2\,-\,(3b-1)p^2\,D_0(p)\,+\,i\e} \,.
\label{6Dpp0}
\eeq
In the limit $m_c\to 0$ the above expression, as opposed
to Eq.~(\ref {sing}), is {\em regular}. 
Similar calculations can be done in the $N>2$ case. The results is
proportional to
\beq
{\,p_\mu \,p_\nu\,D(p,y) \over (2+N) \,u^N\,-\,k_N\,p^2\,D_0(p)\,+\,i\e} \,,
\label{Dpp0}
\eeq
which is also regular in the $m_c\to 0$ limit
where it  approaches the 4D expression.
Therefore, we conclude  that there is no reason to
expect any breaking of  perturbation theory in the model (\ref {1})
 below  the scale of its UV cutoff.

\vspace{1mm}

If $b<1/3$ and $N=2$  we find,  by the same token,
\beq
{p_\mu \,p_\nu\,D(p,y)\over 2} \left (  {
1 \over 2m_c^2\,-\,(3b-1)p^2\,D_0(p)\,+\,i\e} \,+\,
(\e\to -\e) \right )\,.
\label{6Dpp}
\eeq
Again, in the limit $m_c\to 0$ the above expression, in contradistinction with
 Eq.~(\ref {sing}), is regular. Moreover, in this limit
(and at $y=0$) it approaches the 4D expression,
in a particular gauge.
Analogous calculations can be readily done in the $N>2$ and $b<(2N-2)/3N$  
case. The results is
\beq
{p_\mu \,p_\nu\,D(p,y)\over 2} 
\left[
{1 \over (2+N) u^N - k_N\,p^2\,D_0(p)+ i\e}
+ (\e\to -\e)
\right] .
\label{Dpp}
\eeq
This expression is also regular in the $m_c\to 0$ limit
where it  arrives at the correct 4D limit.
We conclude therefore,  that in the general case  there is no reason to
expect any breaking of  perturbation theory in the model (\ref {11})
below  the scale of its UV cutoff. Note that the 
expressions (\ref {6Dpp}) and (\ref {Dpp}) are singular for 
small Euclidean momenta $p^2\sim - m_c^2$. By construction 
this singularity has no 
imaginary part and there is no physical state associated with it. 
One might expect 
that this singularities will be removed after the loop corrections 
are taken into account 
in a full quantum theory. These considerations are beyond the scope 
of the present work.
 
An analogy with the Higgs mechanism for non-Abelian gauge fields
is in order here. For massive non-Abelian gauge fields
nonlinear amplitudes violate the unitarity bounds at the scale set by the 
gauge field mass.
This disaster is cured  through the  introduction of  the Higgs field.
Likewise, nonlinear amplitudes
of the 4D massive gravity (\ref {PF}) blow up
at the scale $\Lambda_m$. The unwanted explosion is canceled 
at the  expense of introducing an infinite number of the
Kaluza--Klein fields in  (\ref {1}).

\subsection{Perturbation theory in nonflat backgrounds}

So far we have been discussing perturbations about a brane 
that has no tension. This was done just to learn what kind of gravity was 
produced by the brane induced term on the world-volume. However, the 
problem of real physical interest is to do the same calculation 
with a brane that has an arbitrary tension. 
For $N=2$ case this was studied
in Ref. \cite {zura2} with the conclusion that the properties
obtained in Refs. \cite {DGP,DG,DGKN1,DGHS} hold unchanged.
In the next two subsections  
we will perform the qualitative analysis for $N>2$ in terms of the Green's 
function on the brane following Refs.  \cite {DG,DGHS},  
as well as in terms of the KK modes following the method
of Ref. \cite {DGKN1}. We will show that at observable distances
the 4D laws of gravity are indeed reproduced.
  
\subsubsection{Propagator analysis}

The nature of gravity on the brane perhaps is simpler understood from
the propagator analysis. The equation for the graviton two-point
Green's function (we omit tensorial structures) 
takes the form
\beq
\m^{2+N}{\hat {\mathcal O}}_{4+N}\,G(x,y)&+&{\mpl^2\,\delta
(y)
\over y^{N-1}\,A^2(y)}\,
{\hat {\mathcal O}}_4\,G (x,0)
\nonumber\\[3mm]
&=& T\,\delta^{(4)}(x)\,{\delta
(y)\over y^{N-1}}\,,
\label{Prop}
\eeq
where $T$ denotes the source (which will be put equal to 1 below) and 
\beq
{\hat {\mathcal O}}_{4+N}\equiv
{1\over \sqrt{G}}\partial_A\sqrt{G}G^{AB}
\partial_B+{\rm higher~derivatives},~~ 
{\hat {\mathcal O}}_4\equiv  \partial^\mu\partial_\mu .
\label{B0}
\eeq
Using the technique of Ref. \cite{DG}, the scalar part of the solution in the
Euclidean four-momentum space can be written as\,\footnote{
Note that in the warped case the scale $M_{\rm ind}$ differs from that 
of the flat case by a constant multiplier $A^2(\Delta)$. For simplicity 
this won't be depicted manifestly below.}
\beq
G(p,y )\,= \,{D(p,y) \over\mpl^2 p^2 D(p, 0) \,+\, \m^{2+N}}\,,
\eeq
where $D(p,y)$ is the Euclidean 4-momentum Green's function of the
bulk operator ${\hat {\mathcal O}}_{4 + N}$, that is 
${\hat {\mathcal O}}_{4 + N} D(p,y)=\delta(y)/y^{N-1}$. 
What is crucial for us is the behavior of the 
Green's function on the brane,
\beq
G(p,y=0 )\,=\,{1 \over \mpl^2 p^2\,+\,\m^{2+N} D^{-1}(p, 0)}\,.
\label{gee}
\eeq
Let us discuss this expression first. The denominator in 
(\ref {gee}) consists of two terms. The first term, 
$\mpl^2 p^2$, is what gives rise to 4D behavior. 
The second term in the denominator, $\m^{2+N} D^{-1}(p, 0)$, sets the 
deviation from the 4D laws and is due to the infinite-volume extra bulk.
Therefore, in the regime when $\mpl^2 p^2$ dominates over 
$\m^{2+N} D^{-1}(p, 0)$ we get 4D laws, while in the opposite case
we obtain the higher-dimensional behavior. The question is what is
the crossover scale at which this transition occurs.
To answer this question we need to know
the expression for $D(p,0)$. Let us start for simplicity
with the case when ${\mathcal E}=0$, i.e.,  
the background metric is flat. 
We will denote the corresponding Green's function 
by $D_0(p, y)$ to distinguish it from $D(p,y)$.
Moreover, let us drop for a moment
higher-derivatives in the expression for 
${\hat {\mathcal O}}_{4 + N}$.  In this case 
$D_0(p,y)$ is nothing but the Green's function of 
the  $(4+N)$-dimensional d'Alambertian. Its behavior at 
the origin is well known,
\beq
D_0(p,y \to 0)\,\sim \,{1\over y^{N-2}}\,.
\label{dal}
\eeq  
Hence, $D_0(p,0)$ diverges and therefore the term 
$\m^{2+N} D^{-1}(p, 0)=\m^{2+N} D_0^{-1}(p, 0) $
in Eq. (\ref {gee}) goes to zero. This would indicate that
4D gravity is reproduced on the brane at all distances. However, 
the UV divergence in (\ref {dal}) is unphysical. 
This divergence is smoothed out by UV physics 
\cite {DG,Wagner,Kiritsis,DGHS}. 
In reality the bulk action and the operator 
${\hat {\mathcal O}}_{4 +N}$ contain an infinite number of
high-derivative terms that should smooth out singularities in the 
Green's function in  (\ref {dal}).
Since these high-derivative operators (HDO) are suppressed by the scale $\m$, it is natural that
the expressions (\ref {dal}) is softened at the very same scale 
$y \sim \m^{-1}$. As a result one obtains \cite {Kiritsis,DGHS}
$D_0(p,y=0)\sim \m^{N-2} (1+{\mathcal O} (p/\m))$. 
Substituting the latter expression
into (\ref {gee}) we find that $$\m^{2+N} D^{-1}(p, 0)= \m^{2+N} 
D_0^{-1}(p, 0)\sim \m^4 \,.$$
Therefore the crossover scale is $r_c\sim \mpl /\m^2 \sim 10^{28}$ cm.
At distances shorter than $\sim 10^{28}$ cm the 4D laws dominate.

Let us now switch on   effects due to a nonvanishing tension ${\mathcal E}$. 
The background in this case is highly distorted. 
The distortion is especially strong near the brane.
Let us start again with the case when the HDO's are neglected and 
${\hat {\mathcal O}}_{4 +N}$ contains only two derivatives at most.
Then, the expression for the $D$-function reduces to
\beq
D(p,y)\,\sim\,D_0(p,y)\,{\mathcal F}(y_g / y )\,,
\label{solsol}
\eeq  
where, as before, $D_0(p,y) \sim {1/y^{N-2}} $  and 
${\mathcal F}$ is some function which is completely determined by 
the background metric (by the functions $A,B$ and $C$)
and ${\mathcal F}(0)={\it const}$.  In the region where the solution 
of the Einstein equations can be trusted, 
${\mathcal F}$ can be approximated as follows,
$ {\mathcal F}(y_g / y)=( {y_g / y} )^{\alpha^2}+c$,
where $\alpha$ and $c$ are some constants determined by $N$.
If we were to trust this solution all the way down to the point 
$y =0$ we would obtain again that  $\m^{2+N} D^{-1}(p, 0)=0$ and 
that gravity is always four-dimensional on the brane.
However, as we discussed above (see also the previous section), 
the  existence of high-derivative terms tells us that the 
background solution cannot be trusted for distances
$y \ll y_*$. 
In general, $y_*=\m^{-1}(y_g \m)^{\gamma}$ with 
$ \gamma \ll 1$ and  $y_* \lsim  y_g$.
Thus, for $y \ll y_*$  the higher curvature 
invariants become large in units of $\m$ and infinite number of them should 
be taken into account. In order to find the effect of this softening,
let us take a closer look at the expression (\ref {solsol}).
There are two sources of singularities in this expression. 
The first one  emerges on the right-hand side of (\ref {solsol})
as a multiplier, $D_0\sim {1/ y^{N-2}}$; this  singularity 
was discussed above in (\ref {dal})
and is independent of the background geometry.
Instead, it emerges when the operator 
${\hat {\mathcal O}}_{4 +N}$ is restricted to the quadratic order only.
We expect that this singularity, as before, is softened at the 
scale $\m^{-1}$ after the higher derivatives are 
introduced in ${\hat {\mathcal O}}_{4 +N}$. Hence, in (\ref {solsol})
when we take the limit $y \to 0$ 
we should make a substitution $$D_0\sim {1/ y^{N-2}} \to \m^{N-2}
(1+{\mathcal O} (p/\m))\,.$$
On the other hand, the second source of singularity in 
(\ref {solsol}) is due to the function  ${\mathcal F}$.
This singularity is directly related to the fact that 
the background solution breaks down at distances of the order of 
$y_*$. As we discussed in the previous section, 
the UV completion of the theory by HDO's should smooth out 
this singularity in the background solution.
In order to get the crossover scale
we can use the following procedure which overestimates
the value of   $\m^{2+N} D^{-1}(p, 0)$.
In the limit $y \to 0$ we could make the   substitution 
$ y_g/y \to y_g/y_*$ in
the expression for ${\mathcal F}$ and in
(\ref {solsol}).
Using these arguments we find $$ D(p, 0)\lsim \m^{N-2} \left[
({y_g / y_*})^{\alpha^2}+c \right].$$ Moreover, taking into account that
$y_* \lsim y_g$ we get $$\m^{2+N} D^{-1}(p, 0)\lsim \m^{4}\,.$$
Therefore, we conclude that, as in the zero-tension case, 
the crossover distance\footnote{If we were to assume  that the background 
solution is softened at $\m^{-1}$ rather than at $y_*$,
we would obtain even larger value for the 
crossover scale. This can be turned around to make  the following 
observation. If the background metric softens at $\m^{-1}$, and/or 
if $y_g\gg y_*$, the value of $\m$ should not necessarily 
be restricted to $10^{-3}$ eV, but  can be much higher. 
Unfortunately, these properties does not seem to allow 
further analytic investigation.} in the nonzero tension  
case can be of the order of $10^{28}$ cm. 
 
\subsubsection{Kaluza--Klein  mode analysis}

The purpose of this section is to study the
effect of a nonvanishing brane tension on
4D gravity in terms of the KK modes.
The Einstein equations (with up to two derivatives)  
that follow from the action (\ref {actD}) can be written as
\beq
&&\m^{2+N}\,\left ({\mathcal R}_{AB}\,-\,{1\over 2}G_{AB}\,{\mathcal R}\right )
\, +\, \mpl^2\,\delta^\mu_A \delta^\nu_B \, \d \,
\left ({ R}_{\mu\nu}\,-\,{1\over 2}{g}
_{\mu\nu}\,{ R} \right )\,\nonumber\\[0.3cm] && =\,
 {\mathcal E} \,{ g}_{\mu\nu}\,\delta^\mu_A \delta^\nu_B \d \,.
\label{eeq1}
\eeq
Below we consider fluctuations $h_{\mu\nu}(x,y_n)$ that are relevant
to 4D interactions on the brane,
\beq
ds^2&=&A^2(y)\,\left [\eta_{\mu\nu}\,+\,h_{\mu\nu}(x,y_n)\right]
\,dx^\mu dx^\nu
\nonumber\\[2mm]
&-& 
B^2(y)\,dy^2\,-
\,C^2(y)\,y^2\,d \Omega^2_{N-1}\,.
\label{fluct1}
\eeq
As  typically happens in warped backgrounds,
equations for graviton fluctuations
are identical to those for a minimally coupled
scalar~\cite {Borut,Csaki}. The present case is no
exception.  Equation (\ref {eeq1}) on the
background defined in Eq.~(\ref {fluct1}) takes the form
\beq
\m^{2+N}{\hat {\mathcal O}}_{4+N}\,h_{\mu\nu}(x,y_n)\,+\,{\mpl^2\,\delta
(y)
\over y^{N-1}\,A^2(y)}\,
{\hat {\mathcal O}}_4\,h_{\mu\nu}(x,0)\,=\,0\,,
\label{BBox}
\eeq
where
\beq
{\hat {\mathcal O}}_{4+N}\,\equiv
\,{1\over \sqrt{G}}\partial_A\,\sqrt{G}\,G^{AB}\,
\partial_B\,,~~~~~~~
{\hat {\mathcal O}}_4\, \equiv  \,\partial^\mu\partial_\mu \,.
\label{B}
\eeq
To simplify  Eq.~(\ref {BBox}) we
turn to spherical coordinates with respect to $y_n, n=1,2,.., N$,
and decompose fluctuations as follows:
\beq
h_{\mu\nu}(x,y)\,\equiv \, \epsilon_{\mu\nu}(x)\,\sigma(y)\,
\phi(\Omega)\,,
\label{fluc}
\eeq
where the components in Eq.~(\ref {fluc}) satisfy
the conditions
\beq
{\hat {\mathcal O}}_{4+N}\,\epsilon_{\mu\nu}(x)\,& = &\,{1\over A^2}\,
\partial^\mu\partial_\mu
\,\epsilon_{\mu\nu}(x)\,=\,-{m^2\over A^2}\,\epsilon_{\mu\nu}(x)\,,
\label{eps} \\[0.3cm]
{\hat {\mathcal O}}_{4+N}\,\phi(\Omega)\,& = &
{l\,(l+N-2)\over C^2\,y^2}\,\phi(\Omega)\,.
\label{phi}
\eeq
Using these expressions we rearrange Eq.~(\ref {BBox})
as follows:
\beq
&&\left \{ {1\over \sqrt{G}}\partial_y\,\sqrt{G}\,G^{y y}\,
\partial_y\,+\,{l\,(l+N-2)\over C^2\,y^2}\right.
\nonumber\\[3mm]
&-&\left.
{m^2\,\mpl^2\,\delta (y)
\over \m^{2+N}\,y^{N-1}\,A^2(y)}
\right \}\,\sigma\,=\,
{m^2\over A^2}\,\sigma\,.
\label{eq}
\eeq
Our goal is to rewrite  this expression
in the form of a Schr\"odinger equation
for fluctuations of mass $m$.
We follow the method of Refs.~\cite {CK2,Emparan}.
It is useful to introduce a new function
\beq
\chi\,=\,{G^{1/4}\over \sqrt{A\,B}}\,\sigma\,,
\label{chi}
\eeq
and a new coordinate
\beq
u\,\equiv\,\int_0^{y}\,d\tau \,{B(\tau)\over A(\tau)}\,.
\label{u}
\eeq
In terms of these variables Eq.~(\ref {BBox})
takes the form
\beq
\left \{
-{d^2\over du^2}+V_{\rm eff}(u)
+{A^2\,l(l+N-2)\over C^2y^2}
-{m^2\mpl^2\,\delta (y)
\over \m^{2+N}\,y^{N-1}}
\right \}\chi=
{m^2}\chi\,,
\label{equ}
\eeq
where the effective potential $V_{\rm eff}(u)$
is defined as
\beq
V_{\rm eff}(u)\,=\,{\sqrt{A\,B}\over G^{1/4} } \,
{d^2\over du^2}\,\left ( {G^{1/4}\over \sqrt{A\,B}}
\right )\,.
\label{Veff}
\eeq
Note that the first two terms in (\ref {equ}) can be
rewritten as 
\beq
-{d^2\over du^2}\,+\,V_{\rm eff}(u) \, = \, \left ({d\over du}\,+\,
 {d {\mathcal B} \over du}  \right )\, \left (-{d\over du}\,+\,
 {d {\mathcal B} \over du}  \right )\,,
\label{QQ}
\eeq
where 
\beq
{\rm exp} \left ( {\mathcal B} \right )\, \equiv\, 
{G^{1/4}\over \sqrt{A\,B}}\,.
\label{B1}
\eeq
With appropriate physical boundary conditions 
the operator on the right-hand side of (\ref {QQ})
is selfadjoint positive-semidefinite with 
a complete set of eigenfunctions of 
nonnegative eigenvalues.

Let us analyze Eq.~(\ref {equ}), in particular, the
properties of the KK modes  following from it.
What is crucial  for our purposes is
the value of the KK wave functions on the brane, i.e.,
$|\chi (m,y=0)|^2$.
The latter determines a potential between two static
sources on the brane \cite {DGKN1}.
We would like to compare the
properties of $|\chi (m,y=0)|^2$ which are known~\cite {DGKN1,DGHS}
 only for ${\mathcal E}\,=\,0$,
with the properties obtained at ${\mathcal E}\ne 0$.

First, we recall  the properties
of $|\chi (m,y=0)|^2$
for $N=1$ and tensionless brane, ${\mathcal E}=0$.
In this case $A=B=1$, $C=0$ and $l=0$. Hence, $u=y$ and
$V_{\rm eff}(u)=0$. Equation (\ref {equ}) becomes
\beq
\left \{
-{d^2\over dy^2}\,
-\,{m^2\,\mpl^2
\over \m^{3}}\, \delta (y)
\right \}\,\chi\,=\,
{m^2}\,\chi\,.
\label{redeq}
\eeq

\vspace{0.1cm}

For each KK mode of mass $m$ there is a delta-function
{\it attractive} potential, the strength of which is proportional
to the mass of the mode itself.
Hence, the higher the mass, the more the influence of the potential is.
The attractive potential leads to a suppression of
the wave function at the origin (suppression of $|\chi (m,y=0)|^2$).
Therefore, the larger the mass of a KK state,
the more  suppressed  is its wave function at zero.

Simple calculations in this case yield
$$|\chi (m,y=0)|^2={4/(4+m^2r_c^2})\,,$$
where $r_c \sim \mpl^2/\m^3$.
This should be contrasted with the expression for $|\chi (m,y=0)|^2$
in a theory with no brane induced term
(i.e., with no potential in Eq.~(\ref {redeq})). In that case 
$|\chi (m,y=0)|^2 \,=\,1$.
We see  that the KK modes with masses $m\gg r_c^{-1}$ are
suppressed on the brane. The laws of gravity on the
brane are provided  by light modes with  $m\,\lsim\, r_c^{-1}$.
This warrants~ \cite {DGP,DGKN1} that at distances
$r\lsim  r_c$ measured along the brane
the gravity laws  are
four-dimensional.

A similar phenomenon takes place for $N\ge 2$, with
a tensionless brane.
Here  $A=B=C=1$, $u=y$ and $V_{\rm eff}(u)=0$.
The Schr\"odinger equation takes the form of Eq.~(\ref {equ})
with the above substitutions.
The total potential consists of an attractive
potential due to the induced term
and a centrifugal repulsive potential.
Because at $y \to 0$ the attractive potential is dominant,
one finds  properties similar to the $N=1$ case.
Heavy KK modes are suppressed on the brane --
at distances $r\lsim  r_c$  the brane-world
gravity is four-dimensional. The only difference~ \cite {DGHS} is that
$r_c\sim \mpl/\m^2$  for $N\ge 2$.

Let us now turn to the discussion of the
case of interest when  ${\mathcal E}\ne 0$
and $A,B,C\ne 1$. Here the complete
equation (\ref {equ}) must be studied.
For the solutions that soften due to the HDO's close to  
the brane core we expect that as $y\to 0$, $u\sim y$.
Hence, to study the suppression of the wave functions
on the brane one can replace $d^2/du^2$ in (\ref {equ}) by
$d^2/dy^2$. The next step is to clarify the role of
the potential $V_{\rm eff}(u)$ that is nonzero when we switch on the
brane tension ${\mathcal E}\ne 0$.
Since a positive tension brane should give rise to an 
additional {\it attractive} potential in space with $N>2$, we expect 
that $V_{\rm eff}(u)$ is negative at the origin (it should tend to 
$-\infty$ at the origin if the  HDO's are not taken into account). 

The warp factors $A,B$ and $C$ contain the only 
dimensionful parameter, $y_g$. So does the potential $V_{\rm eff}(u)$.
Therefore, the maximal value of the potential (if any) in the interval 
$0<y < y_g$ should be  determined by the very 
same scale, ${\rm max}\{V_{\rm eff}\}\sim  y_g^{-2}\,.$

If the form of the attractive potential were trustable all the way down 
to small values of the coordinate, then 
an attractive nature of the potential could  
make easier to obtain 4D  gravity on a brane as   
compared to the zero tension case. Unfortunately we cannot
draw this conclusion since the expression for the potential
is not trustable below the distance scales $y <y_*$
(see discussions in the previous section).
Although $y_*$ is smaller than $y_g$, nevertheless
this two scales can have the same order of magnitude.
Based on the discussions in the previous
section one should expect that the potential in the full theory
softens below $y_*$ and does not really  give rise to a substantial
attraction below that scale.
On the other hand, the potential could  give rise to some 
undesirable results. Indeed, it could produce  
a bump (a potential barrier) at some finite distance
from the core somewhere in the interval $0<y <y_g$.
For a parameter range for which this 
discussion is applicable (i.e., for $y_g^{-1}\ll \m$)
the hight of the bump can be of the order of  
${\rm max}\{V_{\rm eff}\}\sim  y_g^{-2}\,.$
A KK mode with the mass $m \gsim y_g^{-1}$ will not
feel the presence of of such $V_{\rm eff}$.
Its wave function will have the same properties as
in the  tensionless brane theory (i.e.  the modes with
$m > r_c^{-1}$ will be suppressed on the brane).
However, the wave function of any KK mode with the mass  
$m \lsim y_g^{-1}$
will be additionally suppressed on the brane because of  the
potential barrier in $V_{\rm eff}$. The question is whether this
effect can alter the laws of 4D gravity on the brane at observable
distances. If $y_g$ is small this effect will certainly
spoil the emergence of 4D gravity on a brane. 
The reason is that the KK modes that are 
lighter than $y_g^{-1}$ will be 
additionally suppressed on the brane.
If these were  the ``active'' modes that participate
in the mediation of 4D gravity at observable distances
in the tensionless case, then having them additionally 
suppressed would change the 4D laws.    
However, if $y_g$ is sufficiently large the modes which are 
additionally suppressed are very light $m<y_g$, and, if so, 
``switching  off'' these modes won't be important for 4D gravity. 
For instance, if $y_g \gsim 10^{27}$ cm, 
as it happens to be the case
in the present model, gravity at observable distances
will not be noticeably different from gravity on a
tensionless brane.

Therefore, we arrive at the following qualitative
conclusion. In the worst case, gravity on the brane 
world-volume is mediated
by the KK modes that have masses in the band
$ y_g^{-1} \lsim m  \lsim r_c^{-1}\,.$
Hence, at distances $r\lsim y_g$ the effects
of the brane tension are negligible and gravity on a
brane reproduces the known four-dimensional laws.
Moreover, in a simple case when
$y_g^{-1}\sim r_c^{-1}$, one can think,
qualitatively, that gravity on the brane is mediated
by a 4D graviton of  mass $$m_g \sim y_g^{-1} \sim r_c^{-1}\,.$$
In the present context this value is of the order of
the Hubble scale $$m_g \sim H_0 \sim 10^{-33}\,{\rm eV}\,.$$
A graviton with such a small mass is consistent with
observations.

\section{Brief summary} 

We reviewed the arguments why large distance modification of gravity is 
a promising direction toward the solution of   CCP.  
The models that modify gravity at large distances emerge 
in the context of braneworlds with infinite-volume 
extra dimensions. The 4D interaction  in these models 
of brane induced gravity  is due to the world-volume 
Einstein--Hilbert term. We discuss in detail the 
properties of these models in dimensions five and higher.
The 5D model, although not capable of solving  CCP,
is nevertheless a consistent theory of a large-distance modification 
of gravity with many instructive properties. These properties, 
that are scattered in the literature, were collected in Sect. 3
of the present article. Furthermore, we discussed 
brane-induced gravity in dimension six and higher. These 
models share certain properties  with the five-dimensional theory. 
and, at the same time, they
differ from it too. In particular, they could lead to   solution of 
 CCP. To establish whether or not this is the case  one needs to 
 perform further detailed 
calculations the algorithms of which were 
outlined in detail in Sect.~4.

\section*{Acknowledgments}

I would like to thank  Nima Arkani-Hamed, 
Cedric Deffayet, Savas Dimopoulos,  Gia Dvali, Andrei Gruzinov, 
Xin-rui Hou, Alberto Iglesias, Marko Kolanovi\'c, Arthur Lue, 
Francesco Nitti,  Massimo Porrati, Emiliano Sefusatti  and Arkady 
Vainshtein  for fruitful collaboration and discussions on the issues  
of this article. I am especially grateful to 
Misha Shifman for all the above and for his help to improve 
the manuscript.

\newpage
%\begin{thebibliography}{99}
\addcontentsline{toc}{section}{References}
%\bibliographystyle{h-physrev4}
%\bibliography{bibliography}

\end{document}